\documentclass[floatfix,reprint,pra,aps,twocolumn,groupedaddress,amsmath,amssymb,footinbib]{revtex4-2}
\usepackage{xcolor}
\usepackage{graphicx}  % needed for figures
\usepackage{dcolumn}   % needed for some tables
\usepackage{bm}        % for math
\usepackage{verbatim}   % for math
\usepackage[mathscr]{euscript}
\usepackage{epstopdf}
\usepackage{csvsimple}
\usepackage{float}  % needed for positioning of table
\usepackage{ragged2e}
\usepackage{soul}        %JV for strikethrough
\usepackage{marginnote}  %JV for margin notes
\usepackage{multirow}
\pdfoutput=1
\usepackage[normalem]{ulem}

\usepackage[absolute]{textpos}  % for the \margin{}{} command

\usepackage{mathtools}

\newcommand\Mycomb[2][^n]{\prescript{#1\mkern-0.5mu}{}C_{#2}}

\newcommand{\bea}{\begin{eqnarray}}
\newcommand{\eea}{\end{eqnarray}}

\newcommand{\ket}[1]{|#1\rangle}
\newcommand{\breaka}[1]{\langle#1|}
\newcommand{\dn}{{\downarrow}}
\newcommand{\up}{{\uparrow}}
\newcommand{\ip}[1]{{\langle #1 \rangle}}
\newcommand{\eq}[1]{Eq.~(\ref{#1})}
\newcommand{\eqs}[1]{Eqs.~(\ref{#1})}
\newcommand{\eqr}[1]{(\ref{#1})}

\usepackage{hyperref}
\hypersetup{
    colorlinks = true,
    linkcolor=blue
}

\colorlet{Mycolor1}{green!10!orange!90!}
\definecolor{mint}{rgb}{0.85,1,0.85}
\definecolor{dgreen}{rgb}{0,0.2,0}
\definecolor{RED}{rgb}{1,0,0}
\definecolor{JVcolor}{rgb}{1, 0, .75}

\makeatletter
    \newcommand{\jv}{\@ifstar{\jvA}{\textcolor{JVcolor}{(JV)}\jvA}}
    \newcommand{\jvA}[3]{  
        \ifx&#1&\empty\else\textcolor{JVcolor}{[\textit{{#1}}]}\fi    %if #1 is not empty write #1 as comment:  [#1]
        \ifx&#2&\empty\else{\textcolor{JVcolor}{\st{#2}}}\fi        %if #2 is not empty strikeout #2 (i.e. delete)
        \ifx&#3&\empty\else\textcolor{JVcolor}{{#3}}\fi
      }          %is #3 is not empty insert #3 as new text
\makeatother
\newcounter{TxtBlkPos}[page]

\newcommand\rout{\bgroup\markoverwith{\textcolor{red}{\rule[0.5ex]{2pt}{0.8pt}}}\ULon}

\newcommand\routh{\bgroup\markoverwith{\textcolor{blue}{\rule[0.5ex]{2pt}{0.8pt}}}\ULon}
\newcommand{\tc}[2]{{\routh{#1}}{\color{blue}#2}}

\colorlet{Mycolor1}{green!10!orange!90!}

\newcounter{xx}

\renewcommand{\P}{\mathcal{P}}
\pdfoutput=1

\begin{document}

% XXXX delete for final version
% settings for the textpos package used in \margin{}{}
\setlength{\marginparwidth}{50mm}
\setlength{\TPHorizModule}{\marginparwidth}
\setlength{\TPVertModule}{\baselineskip}
\textblockorigin{165mm}{5mm}

\title{Memory erasure with finite-sized spin reservoir}
\author{{Toshio} Croucher}
\affiliation{
   Centre for Quantum Dynamics,\\
   Griffith University,\\
   Brisbane, Queensland 4111, Australia
   }
   \author{Joan A. Vaccaro}
\affiliation{
   Centre for Quantum Dynamics,\\
   Griffith University,\\
   Brisbane, Queensland 4111, Australia
   }
\date{\today}

\begin{abstract}
Landauer's erasure principle puts a fundamental constraint on the amount of work required to erase information using thermal reservoirs. Recently this bound was improved to include corrections for finite-sized thermal reservoirs. In conventional information-erasure schemes, conservation of energy plays a key role with the cost of erasure. However, it has been shown that erasure can be achieved through the manipulation of  spin angular momentum rather than energy, using a reservoir composed of energy-degenerate spin particles under the constraint of the conservation of spin angular momentum, in the limit of an \emph{infinite} number of particles. In this case the erasure cost is in terms of dissipation of spin angular momentum. Here we analyze the erasure of memory using a \emph{finite-sized} spin reservoir. We compute the erasure cost to compare it with its infinite counterpart and determine what size of finite reservoir gives similar erasure cost statistics using the Jensen-Shannon Divergence as the measure of difference. Our findings show that erasure with finite-sized reservoirs results in the erasure of less information compared to the infinite reservoir counterpart when compared on this basis. In addition we discuss the cost of resetting the state of the ancillary spin particles used in the erasure process, and we investigate the degradation in erasure performance when a finite reservoir is repeatedly reused to erase a sequence of memories.
\end{abstract}

\maketitle

\section{Introduction}

Landauer's information erasure principle, first conceived 6 decades ago, still remains a hot research topic today. It places a minimum work cost of $W \geq \beta^{-1} \ln 2$, where $\beta$ is inverse temperature, \cite{Landauer.R1961, Bennett1987} for the erasure of 1 bit of information in memory. Reeb and Wolf \cite{Reeb_2014, Reeb_Wolf2015} derived an improved bound for the heat, $\Delta Q$, dissipated during the erasure of entropy, $\Delta S$, for a finite-sized thermal reservoir given by
\bea
    \beta \Delta Q \geq  \Delta S + \frac{(\Delta S) ^ 2}{\log ^ {2}({d - 1}) + 4} \label{eqn:Improved LB}
\eea
where $d$ is the dimension of the thermal reservoir, which recovers Landauer's original bound in the limit $d \to \infty$. Timpanaro {\it et al.} \cite{Timpanaro2020} derived a tighter bound for arbitrary-sized reservoirs for the zero-temperature limit where Landauer's bound becomes trivial.

Recently Faria and Bonan\c{c}a \cite{Faria2020} numerically verified the finite-bath fluctuation theorem and following this Lobejko \cite{Lobejko2021} derived a tight second law inequality for finite-sized heat baths using ergotropy. In addition, Richens {\it et al.} \cite{Richens2018} derived a finite-bath correction for the second law of thermodynamics. A bound tighter than the Reeb and Wolf bound \eq{eqn:Improved LB} was derived by Goold, Richens {\it et al.} \cite{Goold2015} for erasure of a finite-sized spin reservoir under certain parameters. Spin reservoirs were also used by Pezzutto {\it et al.} \cite{Pezzutto2016} to scrutinize and later validate Landauer's bound. 

Using Jaynes maximum entropy principle \cite{Jaynes1957a,Jaynes1957b} Vaccaro and Barnett \cite{Vaccaro2011,Barnett2013} expanded Landauer's bound to memory erasure of multiple conserved quantities and found
\bea
\Delta V_{k} > v_{k} \ln 2,
\eea
where $V$ is a physical variable indexed by $k$ and $v_k$ represents an associated inverse temperature. For example, if the $z$ component of spin angular momentum is chosen as the conserved quantity i.e. $V_k=J_z$ we find
\bea
    \Delta J_{z} \geq \gamma^{-1}\ln{2}. \label{eqn:VB}
\eea
where $\Delta J_{z}$ is the total (reservoir and memory) change in spin angular momentum and $\gamma^{-1}$ is the spin temperature \cite{Vaccaro2011,Barnett2013}
\bea
      \gamma=\frac{1}{\hbar}\ln\left[\frac{N\hbar-2\langle \hat{J}^{(R)}_{z}\rangle}{N\hbar+2\langle \hat{J}^{(R)}_{z}\rangle}\right]=\frac{1}{\hbar}\ln\left[\frac{1-\alpha}{\alpha}\right].\label{eqn:gamma}
\eea
Here, the superscript (R) represents the reservoir, $\langle\hat{J}^{(R)}_z\rangle=\left(\alpha-\frac{1}{2}\right)N\hbar$ is the $z$ component of the reservoir spin angular momentum, $N$ is the number of spins in the reservoir and $\alpha$ represents the spin polarisation parameter bounded such that $0 \le \alpha \le 1$.

Vaccaro and Barnett presumed the spin reservoir contained a very large number of spins to ensure that its spin temperature remained approximately constant. The requirement that temperature is constant throughout the erasure protocol is a common condition in conventional treatments involving thermal reservoirs. Here we will analyze what happens when this condition is relaxed for a finite-sized spin reservoir. The key difference between the previously discussed research and the research presented here is that the cost of the erasure protocol in our research will be solely in terms of spin angular momentum and not energy.

Accompanying these developments have been advances in the associated resource theories \cite{Guryanova2016, Yunger2016,Halpern2016, Lostaglio2017, Yunger2018, Lostaglio2019, Manzano2020, biswas2022}. Engines that use multiple conserved quantities, more specifically interacting spin and thermal reservoirs has been proposed \cite{Croucher2018, Wright2018}. Research on generalized heat engines with finite sized baths of arbitrary conserved quantities has been analyzed by Ito and Hayashi \cite{Kosuke2018} with the main results showing how to do both identical and identically distribution (i.i.d.)-based scaling and generic scaling with multiple conserved quantities. 

In this paper we extend the analysis of the erasure protocol introduced by Vaccaro and Barnett \cite{Vaccaro2011,Barnett2013}, taking into account pragmatic issues such as finite-sized reservoirs and the cost of returning ancilla spins to their initial state.  We begin by reviewing the Vaccaro-Barnett erasure protocol in \S\ref{sec:review}. Whereas Vaccaro and Barnett presumed the spin reservoir contained a very large number of spins to ensure that its spin temperature remained approximately constant throughout the erasure process, we investigate the modifications to the protocol needed to accommodate a finite sized reservoir in \S{\ref{sec:finite reservoir}. In \S{\ref{sec:ancilla retrival}} we analyze the non-trivial cost of returning  ancilla  spins  to  their  initial  state. We explore the repeated use of the same finite reservoir to erase additional memory spins showing how the reservoir's erasure ability reduces each time. Finally we summarize the  major results within the paper in \S{\ref{sec:conclusion}}.

\section{Model Review \label{sec:review}}

We assume that all degrees of freedom other than the spin degree of freedom do not play an active role in the erasure process, and hence they are traced over in our analysis. In addition, the states of the model are energy degenerate allowing us to focus solely on spin and ensures that the erasure process incurs no energy cost.  The erasure protocol entails performing operations on a memory, the contents of which we would like to erase, using a spin reservoir to store entropy erased from the memory, and spin ancillas to aid in the process. The memory is a spin-$\frac{1}{2}$ particle which we describe exclusively using the basis given by eigenstates of the z component of spin angular momentum. The  memory has a probability of $p_\uparrow$ being in the spin-up state $\ket{\up}\breaka{\up}$ (logical 1) and $p_\downarrow$ in the spin-down state $\ket{\dn}\breaka{\dn}$ (logical 0). This probability will evolve through the erasure process and will go from $p_\uparrow = p_\downarrow = 0.5$ initially to  $p_\uparrow = 0$ and $p_\downarrow = 1$ for the memory to be fully erased.

The spin reservoir consists of $N$ spin-$\frac{1}{2}$ particles in spin equilibrium at inverse spin temperature $\gamma$ and will store the entropy removed from the memory particle. The spin reservoir is presumed to be initially in an equilibrium state described by the probability distribution
\bea
P_{\uparrow}(n,\nu)=\frac{e^{-\gamma n\hbar}}{Z_R},\qquad Z_R=(1+e^{-\gamma\hbar})^N \label{eqn:probr}
\eea
for $0\le n\le N$\tc{}{,} where $n$ is the number of particles in the spin-up state,  $\nu=1, 2,\ldots {}^{N}C_{n}$ indexes different states with the same value of $n$ and $Z_{R}$ is the partition function. From here onward we restrict $\alpha$ to $0 \le \alpha \le 0.5$ as this provides us with positive values of the inverse spin temperature $\gamma$.

The erasure protocol starts with the memory system being brought into spin-exchange contact with the reservoir and allowed to reach equilibrium in their spin degree of freedom. The spin-exchange contact is assumed to conserve the total spin angular momentum while allowing the reservoir and memory systems to exchange the $z$ component of spin-angular momentum in units of $\hbar$. The process of the reservoir and memory systems reaching equilibrium will be called the \emph{equilibration step}.

After the first equilibration step we require the use of an energy-degenerate ancillary spin-$\frac{1}{2}$ particle which is to be used in conjunction with the memory. This ancilla particle is initially in the state $\ket{\dn}\breaka{\dn}$ corresponding to the logical 0 state. The ancilla particle allows the spin angular momentum of the combined memory and ancilla system to be manipulated to assist with the transfer of entropy from the memory the reservoir. All ancilla particles are returned to their initial state by the end of the erasure protocol to assist in the erasure of another memory system, and so they act as a catalysis. Once the first ancilla is added to the memory, a controlled-not (CNOT) operation is applied to the memory-ancilla system in such a way that the memory acts as the control and the ancilla as the target. The combined process of adding an ancilla and performing the CNOT operation on the memory-ancilla system will be called simply a \textit{CNOT step}. The CNOT operation effectively increases the splitting between eigenvalues of the $z$ component of spin angular momentum  of the two physical states that represent the $0$, $1$ logical states of the memory by $1\hbar$, and leaves the memory-ancilla probabilities unchanged with $p_\up$ for the spin up (logical 1) probability  and $p_\dn = 1 - p_\up$ the spin down (logical 0) probability.

After the CNOT step is applied another equilibration step follows exchanging spin-angular momentum between the reservoir and the combined memory-ancilla system in units of $2\hbar$. The combination of a CNOT step followed by an equilibration step constitutes an \emph{erasure cycle}. Each time an erasure cycle is performed, an additional ancillary spin-$\frac{1}{2}$ particle in the state $\ket{\dn}\breaka{\dn}$ is added to the memory-ancilla system and is the target of another CNOT operation in the CNOT step. This increases the eigenvalue gap of the memory-ancilla system by $1\hbar$ and the subsequent equilibration step is adjusted to exchange spin-angular momentum between the reservoir and the combined memory-ancilla system in corresponding units. In the case of the $m$-th CNOT step the gap increases to $(m+1)\hbar$ with the equilibration step exchanging $(m+1)\hbar$. Erasure cycles are repeated until a desired degree of erasure is reached.

\begin{figure*}
	\centering
	\includegraphics[width=\textwidth]{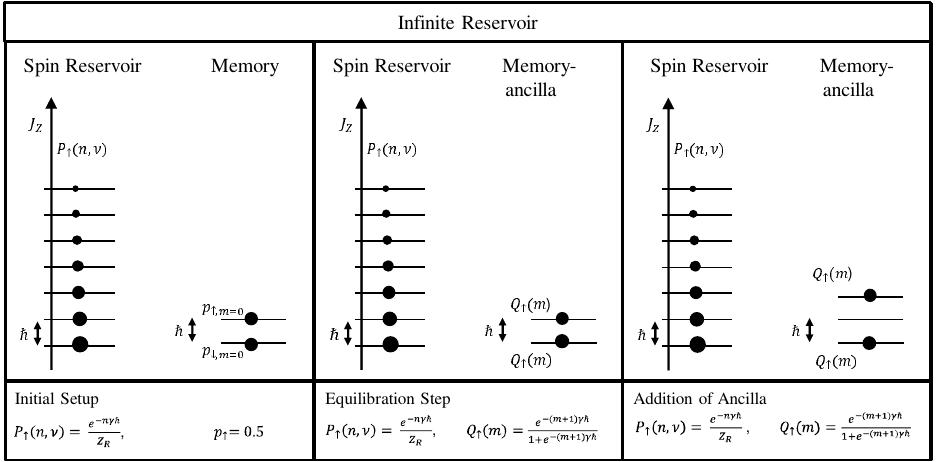}
	\vspace{-2mm}
	%\captionsetup{justification=raggedright,singlelinecheck=false}
	\caption{An illustration of the erasure protocol for an infinite reservoir. The upwards vertical direction represents increasing values of the $z$ component of spin angular momentum. The state of the spin reservoir is represented on the far left by a spin level diagram. The remaining spin level diagrams to its right represent the spin state of the memory-ancilla system. The value of $m$ is the number of CNOT steps that have taken place.}
	\label{fig:infinite}
	\vspace{-2mm}
\end{figure*}

The direct cost of performing the erasure is called the \emph{spinlabor} cost where spinlabor, $\mathcal{L}_{\rm s}$, is the spin-equivalent of work \cite{Croucher2017, Croucher2021}. The spinlabor cost for an erasure cycle is incurred during the application of the CNOT operation. The cost of performing each CNOT operation is $1\hbar$ when the memory-ancilla system is in the spin-up state $\ket{\up}\breaka{\up}$ and zero when it is in the spin-down state $\ket{\dn}\breaka{\dn}$. The probability of the memory being in the spin-up state $\ket{\up}\breaka{\up}$ at the end of an erasure cycle is expressed as
\bea
      Q_\uparrow(m) =\frac{e^{-(m+1)\gamma\hbar}}{1+e^{-(m+1)\gamma\hbar}}  \label{eqn:Q}
\eea
where $m$ is the number of prior CNOT operations that have been applied, one for each completed erasure cycle. Using \eq{eqn:Q} we calculate the expectation value of the total spinlabor cost as 
\bea
\langle\mathcal{L}_{\rm s}\rangle = \sum_{m=1}^{\infty} \hbar Q_\uparrow(m-1), \label{eqn:avg_inf}
\eea
for completely erasing the contents of the memory. This expectation value is bounded such that \cite{Croucher2021}
\bea
  \langle\mathcal{L}_{\rm s}\rangle   & \geq & \gamma^{-1} \ln(1+e^{-\gamma \hbar}).
    \label{eqn:Spinlabor_ineq}
\eea

The result \eq{eqn:avg_inf} is the average over many runs of the erasure protocol. The actual spinlabor cost for each run fluctuates about the average $\langle\mathcal{L}_{\rm s}\rangle$ \cite{Croucher2017,Croucher2018,Croucher2021}. These fluctuations are described by the spinlabor probability distribution $\P^{\text{inf}}_{m}(n)$ which is the probability that the spinlabor cost accumulated after $m$ CNOT operations is $n\hbar$. Here ``inf'' labels the result as being for a reservoir containing an unlimited number of spin particles, i.e. $N\to\infty$ in \eq{eqn:probr}, hereafter referred to as an \emph{infinite reservoir}. $\P^{\text{inf}}_{m}(n)$ is calculated by noting that it is the sum of the probabilities for the two mutually exclusive events possible at the $m$-th CNOT operation: either (i) the accumulated spinlabor cost is $n\hbar$ after the $(m-1)$-th CNOT operation and it does not increase at the $m$-th CNOT operation, or (ii) the accumulated spinlabor cost is $(n-1)\hbar$ after the $(m-1)$-th CNOT operation and it increases by $1\hbar$ at the $m$-th CNOT operation. The corresponding probabilities of each event are   (i) $\P^{\text{inf}}_{m-1}(n)Q_{\dn}(m-1)$ and (ii) $\P^{\text{inf}}_{m-1}(n-1)Q_{\up}(m-1)$). This yields the recurrence relation \cite{Croucher2021}
\bea
    \P^{\text{inf}}_{m}(n)&=&\P^{\text{inf}}_{m-1}(n)Q_{\dn} (m-1)\nonumber\\&&\quad +\P^{\text{inf}}_{m-1}(n-1)Q_{\up} (m-1), \label{eqn:recrel_inf}
\eea
where $Q_\downarrow(m) = 1 - Q_\uparrow(m)$. The spinlabor cost cannot be more than $m\hbar$ where $m$ is the number of erasure cycles completed, hence we restrict $n$ such that $0 \leq n \leq m$. The initial values of $\P^{\text{inf}}_{m=0}(n)$ before any CNOT operations have been applied is
\bea
\P^{\text{inf}}_{0}(n) = \begin{cases} 1, \text{ for }n=0\\
                          0, \text{ otherwise},
            \end{cases} \label{eqn:secondeq m=0}
\eea
since there is no spinlabor cost at that point in the erasure protocol.
We have previously found the analytical solution to \eq{eqn:recrel_inf} in the limit $m\to\infty$ as \cite{Croucher2021}
\begin{align}   \label{eqn:P^inf_infty}
    \P^{\text{inf}}_{\infty}(n)=\frac{ e^{-\frac{1}{2}n(n+1)\gamma\hbar}}{(e^{-\gamma\hbar};e^{-\gamma\hbar})_n(-e^{-\gamma \hbar};e^{-\gamma \hbar})_\infty}
\end{align}
where $(a;q)_n$ is the $q$-Pochhammer symbol \cite{Alvarez-Nodarse1999}
\begin{align}    \label{eqn:ap pochhammer}
        (a;q)_n\equiv\prod_{k=0}^{n-1}(1-a q^k), \quad (a;q)_0\equiv 1.
\end{align}

We note that an alternative protocol that begins with CNOT steps first instead of an equilibration step will have a different total cost of erasure.  The average cost of applying a number of CNOT steps before the first equilibration step is $\frac{1}{2}\hbar$ per operation as $p_\up=\frac{1}{2}$ initially. This cost is unrelated to the spin temperature of the reservoir whereas the average spinlabor cost of all subsequent CNOT steps that occur after the first equilibration step depend on the spin temperature of the reservoir. In particular, at the end of an equilibration step that exchanges spin angular momentum in units of $(m+1)\hbar$, the probability of the memory-ancilla system to be in the spin-up state reduces to $Q_\up(m)$ given in \eq{eqn:Q}, and so the spinlabor cost of the next CNOT step is $Q_\up(m)\hbar$ which is less than $\frac{1}{2}\hbar$. Hence the lowest spinlabor cost occurs when the equilibration step is first applied. We refer to erasure from the first equilibration step as passive erasure (no direct spinlabor cost) and erasure that is driven by erasure cycles as active erasure (costly in terms of directly applied spinlabor) \cite{Croucher2021}. As these alternative CNOT-first versions of the erasure protocol are less efficient, we won't consider them further here. Their intricacies are discussed in \cite{Croucher2021}.

\section{Finite Reservoir  \label{sec:finite reservoir}}
In previous work \cite{Vaccaro2011, Barnett2013, Croucher2017, Croucher2021} we adopted the protocol reviewed in \S\ref{sec:review} which assumes that the reservoir is infinitely large (i.e. $N \rightarrow \infty$) and so the value of $\gamma$ in \eq{eqn:probr} remains constant during the entire erasure process. Although this allows analytical results to be calculated, an infinite reservoir is not realistic. Here, we investigate the more practical situation for performing information erasure using a reservoir containing a finite number of spin particles, which we shall call a \emph{finite reservoir}. 

The use of a finite reservoir immediately raises the following issues and questions. As a finite reservoir has a finite capacity for storing unwanted entropy, it will presumably have lower efficacy for information erasure compared to its infinite counterpart.  How does the spinlabor costs associated with using each type of reservoir compare? Presumably the efficacy of a finite reservoir approaches that of an infinite reservoir as the number of spin particles it contains increases: how large does a finite reservoir need to be to perform equivalently to an infinite reservoir up to a given tolerance? In principle, an infinite reservoir does not change as it is used whereas, due to its accumulation of erased entropy, the spin temperature of a finite reservoir is expected to increase during each run of the erasure protocol; how does a finite reservoir degrade as it is reused?  Also, when using an infinite reservoir, the ancilla are left, in principle, in their initial spin-down state at the end of the erasure protocol: are there additional spinlabor costs associated with returning the ancilla to their initial state when using a finite reservoir? We address these issues and questions in the remainder of this paper.

\subsection{Finite Reservoir Formalism}

%\margin{14}{\jv{\rm (A) I note that $\P^{\text{inf}}_{m}(n)$ in \eq{eqn:recrel_inf}  is the probability of the spinlabor cost after $m$-th CNOT step. It's good to keep the relevance of $m$ in $p_{\up, m}$, $P_{m}(n,M,\nu)$ etc. in this section on ``Finite Reservoir Formalism''  also being after the $m$-th CNOT step.  But it also presents a problem because later we need to calculate probabilities at the beginning and end of an equilibration step following the $m$-th CNOT step.  It seems that $P_{m}(n,M)$ is actually the probability after the $m$-th CNOT step and before the $m$-th equilibration step.\\[2mm]
%In the $\P^{\text{inf}}_{m}(n)$ case, there is no need to mention whether the probability is before or after the equilibration step, but it is quite important to be specific for $P_{m}(n,M)$.\\[2mm]
%We need to say that the CNOT steps change the macrostates without changing the associated probabilities.  Thus after $m$-th CNOT step, the probability of being in the macrostate $\rho(n,M)$ before the $m$-th equilibration step is $P_{m}(n,M)$ and after the $m$-th equilibration step it is $P_{m+1}(n,M)$, i.e. it is the probability for the next CNOT step.\\
%I've done this in comment (B) on next page.
%}{}{}}

%\jv{(A)}{}{}
We begin by modifying the erasure formalism to accommodate a finite-sized reservoir. In particular, we now need to allow for both the reservoir's and memory-ancilla's spin angular momentum polarization (i.e. the probability that each spin particle is in the spin-up state) changing during each equilibration step. Just as for the infinite reservoir case, the initial probabilities for the memory are $p_{\up}=p_\dn=0.5$ and the spin polarisation of the reservoir is $\alpha$ with corresponding inverse spin temperature $\gamma$ given by \eq{eqn:gamma} The initial probability distribution describing the reservoir is therefore given by \eq{eqn:probr} with the value of $N$ fixed. Here, we define $\alpha_{m=0} = \alpha$ as the initial spin polarisation of the reservoir, where $m$ is the number of completed erasure cycles (and, thus, also the number of CNOT steps that have been applied). In turn, the initial spin polarisation of the memory is represented as $p_{\up,m=0}$, $p_{\dn,m=0}$. After the first equilibration step these quantities take on new values with $\alpha_{m=1} = p_{\uparrow,m=1}$, since the memory and reservoir now have the same spin temperature. Since the spin temperature of the reservoir is not constant, as it was for the infinite reservoir case, a new formalism will need to be used.

An eigenstate of the $z$ component of spin angular momentum of the combined reservoir-memory-ancilla system will be written as $|n,M,\nu \rangle_m$ and the corresponding probability of being occupied as $P_{m}(n,M,\nu)$. Here $n$ is the number of spins in the spin-up state for the reservoir, $M$ represents the state of the memory spin with values $0$ or $1$ corresponding to spin-down and -up, respectively, and $\nu$ indexes one of the spin-angular-momentum-degenerate states of the reservoir with $n$ particles in the spin-up state. Each of the states $|n,M,\nu \rangle_m$ represents a microstate of the combined reservoir-memory-ancilla system. Treating the degeneracy index $\nu$ as an internal variable implies that the macrostates are given by the set of microstates $\rho_m(n,M) = \{\ket{n,M,\nu}_m:\nu=1,2,\ldots, {}^{N}C_{n}\}$. The probability of being in the macrostate $\rho_m(n,M)$ is 
\begin{align}   
    P_m(n,M) \equiv \sum_\nu P_m(n,M,\nu). \label{eqn:sum_degen}
\end{align}
Note that the CNOT steps change the states by incrementing the index $m$ without changing the associated probabilities, resulting in the memory-ancilla system becoming out of equilibrium with the reservoir. This disequilibrium is corrected in the subsequent equilibration step at the end of the erasure cycle. For definiteness, we let $P_{m}(n,M)$ in the remainder of this paper represent the joint probability distribution \emph{after} an equilibration step, and  therefore it represents an equilibrium distribution.  The $m$-th erasure cycle is described statistically by}
\bea
  P_{m-1}(n,M) = \tilde{P}_{m}(n,M) \stackrel{\rm eq}{\longrightarrow} P_{m}(n,M)
  \label{eqn:eq_relationship}
\eea
where $\tilde{P}_{m}(n,M)$ represents the disequilibrium distribution of the system being in the macrostate $\rho_m(n,M)$ following the CNOT step, and ``eq'' labels the process representing the equilibration step.

Stepping through the erasure process results in the following scenario. Initially the memory has the probabilities of $p_{\up}=p_{\dn}=0.5$ of being spin-up and spin-down, and the spin polarisation of the $N$-spin  reservoir is $\alpha$. At this point $m=0$, since no erasure cycles have been completed and no CNOT steps have been applied. After the first equilibration step, the memory probabilities of spin-up, -down and reservoir spin polarisation change to $p_{\up,0}$, $p_{\dn,0}$ and $\alpha_{0}$, respectively, and the joint reservoir-memory probability distribution becomes $P_{0}(n,M)$. The first CNOT step costs $1\hbar$ in spinlabor with probability $p_{\up,0}$. This CNOT step brings the reservoir and memory-ancilla system out of equilibrium which is described by the joint probability distribution $\tilde{P}_{1}(n,M)$. An equilibration step follows with the memory probabilities of spin-up, -down and the reservoir spin polarisation changing to $p_{\up,1}$, $p_{\dn,1}$ and $\alpha_{1}$, respectively, and the reservoir-memory-ancilla system returning to equilibrium described by joint probability distribution $P_{1}(n,M)$. The first erasure cycle is now complete. The erasure cycles are repeated until desired with the maximum number of erasure cycles possible being $m=N-1$ since the size of the memory-ancilla system ($m+1$) can not be greater than the reservoir size $N$.

The equilibration process after the $m$-th CNOT step is assumed to randomly exchange spin angular momentum between the reservoir and the memory-ancilla system in units of $(m+1)\hbar$ while conserving the total spin angular momentum. This redistributes the $z$ component of spin angular momentum between the reservoir and memory-ancilla systems making all accessible microstates equally likely, analogously to Boltzmann’s postulate that all microstates of the same energy have the same probability \cite{Tolman1979, Chandler1987, Huang1963}. The accessible states here are those that have the same total $z$ component of spin angular momentum as the combined reservoir-memory-ancilla system before the equilibration step, but allowing for the possibility of the reservoir and memory-ancilla system to exchange either $0$ or $(m+1)\hbar$ of their $z$ component of spin angular momentum. In principle, many exchanges take place in the equilibration process, each one creating entanglement between the reservoir and memory-ancilla systems.  However, the average over all exchanges washes out the entanglement leaving the reservoir and memory-ancilla systems in statistical mixtures of separate spin-angular momentum eigenstates. We therefore concentrate on the net exchange between the reservoir and memory-ancilla systems for the overall equilibration process. Each possibility for net spin exchange is associated with a particular path that the combined reservoir-memory-ancilla system can take. For example, the memory-ancilla system can be left in the spin-down state $\ket{n,0,\nu'}_m$ through the exchange of $0$ spin angular momentum, i.e. by
\bea
|n,0,\nu \rangle_m \rightarrow |n,0,\nu' \rangle_m, \label{eqn:stay_same_1} 
\eea
where the arrow represents the equilibration process and the initial and final degeneracy indices $\nu$ and $\nu'$, respectively, are independent and have values in the ranges
\bea  
  \nu = 1, 2,\ldots {}^{N}C_{n},\quad  \nu' = 1, 2,\ldots {}^{N}C_{n}.\nonumber
\eea
Alternatively, the same final state can be reached through the memory-ancilla system losing $(m+1)\hbar$ of spin angular momentum to the reservoir as in
\bea
|n-m-1,1,\nu \rangle_m    \rightarrow |n,0,\nu' \rangle_m  \label{eqn:switch_1}
\eea
with
\bea
    \nu=1, 2,\ldots {}^{N}C_{n-m-1},\quad  \nu' = 1, 2,\ldots {}^{N}C_{n}.\nonumber 
\eea
Thus, the total set of paths that end in the state $\ket{n,0,\nu'}_m$ after equilibration is given by
  \begin{equation}
  \left.
    \begin{array}{@{}lr@{}}
        |n,0,\nu \rangle_m \\
        |n-m-1,1,\nu \rangle_m       \end{array} \right\} \rightarrow |n,0,\nu' \rangle_m . \label{eqn:paths_1}
  \end{equation}

Similarly, the memory-ancilla system can be left in the spin-up state $|n,1,\nu' \rangle_m$ either through the exchange of $0$ spin angular momentum, i.e. by 
\bea
|n,1,\nu \rangle_m    \rightarrow |n,1,\nu' \rangle_m  \label{eqn:stay_same_2} 
\eea
with 
\bea
    \nu = 1, 2,\ldots {}^{N}C_{n},\quad  \nu' = 1, 2,\ldots {}^{N}C_{n},\nonumber
\eea
or through gaining $(m+1)\hbar$ of spin angular momentum from the reservoir, i.e. by
\bea
|n+m+1,0,\nu \rangle_m    \rightarrow |n,1,\nu' \rangle_m  \label{eqn:switch_2}
\eea
with 
\bea
\nu=1, 2,\ldots {}^{N}C_{n+m+1},\quad \nu'=1, 2,\ldots {}^{N}C_{n}. \nonumber
\eea
Again, summarizing the paths that end in the state $\ket{n,1,\nu'}_m$ after equilibration gives
  \begin{equation}
  \left.
    \begin{array}{@{}lr@{}}
        |n,1,\nu \rangle_m \\
        |n+m+1,0,\nu \rangle_m       \end{array} \right\} \rightarrow |n,1,\nu' \rangle_m . \label{eqn:paths_2}
  \end{equation}

We define $T_{m}(n,M)$ to be the probability that the reservoir-memory-ancilla system transitions to the macrostate $\rho_m(n,M)$ from all accessible prior macrostates during the equilibration step following the $m$-th CNOT step. We know from \eq{eqn:paths_1} that during this equilibration step, the state $\rho_m(n,0)$ can be reached from either of the macrostates $\rho_m(n,0)$ or $\rho_m(n-m-1,1)$.
From the definition of the probability of the event $E$,
\bea
P(E) = \frac{\mbox{\# of outcomes favorable to E}}{\mbox{\# of outcomes in total}},
\eea
and all outcomes are equally likely according to Boltzmann's postulate. The corresponding transition probability is
\bea
T_{m}(n,0) \equiv \frac{^{N}C_{n}}{^{N}C_{n-m-1}+^{N}C_{n}} \label{eqn:T0}
\eea
for $n \ge m+1$ and $T_{m}(n,0)\equiv 1$ otherwise. Similarly, we find the probability for the transitions described by \eq{eqn:paths_2} to be
\bea
T_{m}(n,1) \equiv \frac{^{N}C_{n}}{^{N}C_{n+m+1}+^{N}C_{n}}, \label{eqn:T1}
\eea
%The combinations are defined such that
%\bea
%^{k}C_{j} = 0 \; \; \; \; \; \mbox{if} \; \; \; \; \; j<0 \; \; \; \; \; \mbox{or} \; \; \; \; \;j>k.
%\eea
for $n \le N-m-1$ and $T_{m}(n,0)\equiv 1$ otherwise.

We use \eq{eqn:T0} to calculate the probability $P_{m}(n,0)$ of the system taking the paths in {\eq{eqn:paths_1}} and ending in the macrostate $\rho_{m}(n,0)$, as follows. Let $A$ represent the event that the system ends the equilibration step of the $m$-th cycle in the macrostate $\rho_{m}(n,0)$ and $B$ represent the event that it begins in either of the macrostates $\rho_{m}(n,0)$ or $\rho_{m}(n-m-1,1)$. Using the definition of conditional probability 
\bea
P(A \cap B) = P(A|B ) P(B), \label{eq:cond_prob}
\eea
and interpreting $P(A \cap B)$ as the probability $P_{m}(n,0)$  that we want, $P(A|B)$ as the transition probability $T_{m}(n,0)$, and $P(B)$ as the probability $\tilde{P}_{m}(n,0) + \tilde{P}_{m}(n-m-1,1)$ which, according to \eq{eqn:eq_relationship} is equal to ${P}_{m-1}(n,0) + {P}_{m-1}(n-m-1,1)$, then gives
\bea
P_{m}(n,0) &=&  T_{m}(n,0) [P_{m-1}(n,0) + \nonumber\\
    &&P_{m-1}(n-m-1,1)] \label{eqn:res_mem_prob_down}
\eea
for $0 < m \le N-1$.
The situation for $m=0$ is special as event $B$ then represents the system being in either of the macrostates $\rho_0(n,0)$ or $\rho_0(n-1,1)$ \emph{before} the first equilibration step, i.e. at the beginning of the erasure process. In that case, instead of \eq{eqn:res_mem_prob_down}, we have
\bea
P_{0}(n,0) &=&  T_{0}(n,0) [P_\up(n)p_\dn + P_\up(n-1)p_\up ]. \label{eqn:res_mem_prob_down m=0}
\eea
where we have defined the total initial probability for the reservoir
\bea   
  P_\up(n) \equiv \sum_{\nu=0}^{^{N}C_{n}}P_\up(n,\nu) \label{eqn:P_up(n)}
\eea
and $P_\up(n,\nu)$ is given by \eq{eqn:probr}. Following similar logic we also find
\bea
P_{m}(n,1) &=& T_{m}(n,1) [P_{m-1}(n,1) + \nonumber \\ 
&& P_{m-1}(n+m+1,0)] \label{eqn:res_mem_prob_up}
\eea
for $0 < m \le N-1$ and 
\bea
P_{0}(n,1) &=& T_{0}(n,1) [P_\up(n)p_\up +  P_\up(n+1)p_\dn]. \label{eqn:res_mem_prob_up m=0}
\eea
For convenience, we have defined $P_{m}(n,0)=P_{m}(n,1)=0$ for $n$ outside the range $0\le n \le N$.
An alternate expression for $P_{m}(n,1)$ is given by
\bea
P_{m}(n,1)  &=& \prod_{d=0}^{m}T_{d}(n,1)  \Bigg(P_\up(n)p_\up +  P_\up(n+1)p_\dn + \Bigg. \nonumber \\
&& \Bigg. \sum_{d=1}^{m} \frac{P_{d-1}(n+d+1,0)}{{\prod_{k=0}^{d-1}T_{k}(n,1)}} \Bigg),\label{eqn:other_form}
\eea
is derived in Appendix \ref{sec:rec_sol}. The derivation of the corresponding probability $P_{m}(n,0)$, is straightforward to calculate and is left as an exercise. From $P_{m}(n,M)$ we can obtain the probability of spinlabor cost $\mathcal{L}_{\rm s}$ for a finite reservoir. Remembering that spinlabor is incurred when the memory is in the spin up state $\ket{\up}\breaka{\up}$ ($M=1$) with corresponding probability given by the marginal of the joint distribution $P_{m}(n,1)$ for the memory-ancilla system, i.e.  
\bea
    p_{\up, m} = \sum_{n}^{N} P_{m}(n,1). \label{eqn:memory_prob}
\eea

\begin{figure*}
	\centering
	\includegraphics[width=\textwidth]{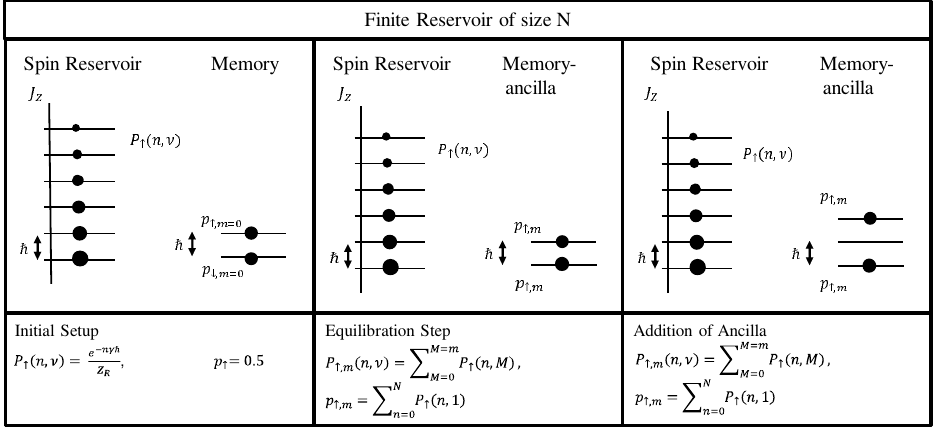}
	\vspace{-2mm}
	%\captionsetup{justification=raggedright,singlelinecheck=false}
	\caption{An illustration of the erasure process for an arbitrary protocol. The upwards vertical direction represents increasing values of the $z$ component of spin angular momentum. The state of the spin reservoir is represented on the far left of each section by a spin level diagram. The remaining spin level diagrams to its right represent the spin state of the memory-ancilla system. The value of $m$ is the number of CNOT steps that have taken place.}
	\label{fig:finite}
	\vspace{-2mm}
\end{figure*}

The total average spinlabor cost is 
\bea
   \langle\mathcal{L}_{\rm s}\rangle = \sum_{m=1}^{N-1} \hbar p_{\up, m-1}, \label{eqn:avg_fin}
\eea
where we have set the upper summation limit to the number of erasure cycle to be one less than the reservoir size $N-1$, since it is not possible to carry out the equilibration step with the memory-ancilla system having an eigenvalue gap size of $(m+1)\hbar$ that is larger than the spread in eigenvalues of the reservoir of $N\hbar$. Substituting \eqs{eqn:other_form} and \eqr{eqn:memory_prob} into \eq{eqn:avg_fin} we find
\bea
\langle\mathcal{L}_{\rm s}\rangle &=&  \hbar \sum_{m=1}^{N-1} \sum_{n}^{N}  \prod_{d=0}^{m}T_{d}(n,1) \Bigg(P_\up(n)p_\up + \Bigg. \nonumber \\
&& P_\up(n+1)p_\dn +  \Bigg. \sum_{d=1}^{m} \frac{P_{d-1}(n+d+1,0)}{{\prod_{k=0}^{d-1}T_{k}(n,1)}} \Bigg). \label{eqn:avg_fin_other_form}
\eea

We obtain a bound for the total average spinlabor cost by considering only the first term of the summation \eq{eqn:avg_fin} as (see Appendix \ref{sec:av spin labor bound} for details)
\bea
\langle\mathcal{L}_{\rm s}\rangle \geq \frac{\hbar N e^{-\gamma \hbar}}{(N+1)(1+e^{-\gamma \hbar})} + \frac{\hbar}{2(N+1)}, \label{eqn:bound_init}
\eea
with the bound being tightest when the reservoir size is small. (Notice that as $N \rightarrow \infty$ the bound approaches the corresponding bound for the infinite reservoir $\langle\mathcal{L}_{\rm s}\rangle \geq \frac{e^{-\gamma \hbar}}{1+e^{-\gamma \hbar}}$.) Similar to the situation for an infinite reservoir, the spinlabor probability distribution $\P^{\text{fin}}_{m}(n)$ for a finite reservoir is 
\bea
    \P^{\text{fin}}_{m}(n)&=&\P^{\text{fin}}_{m-1}(n)p_{\downarrow, m-1}\nonumber\\&&\quad +\P^{\text{fin}}_{m-1}(n-1)p_{\up, m-1}, \label{eqn:recrel_fin}
\eea
where $p_{\up, m}$ is given by \eq{eqn:memory_prob}, the label ``fin'' represents the spinlabor probability for a finite reservoir, and with the initial condition  the same as for the infinite reservoir, i.e.
\bea
\P^{\text{fin}}_{0}(n) = \begin{cases} 1, \text{ for }n=0\\
                          0, \text{ otherwise}.
            \end{cases} \label{eqn:P fin 0}
\eea

\subsection{Spinlabor Cost Comparison \label{sec:Spinlabor Cost Comparison}}
\begin{figure}
	\centering
	\includegraphics[width=0.5\textwidth]{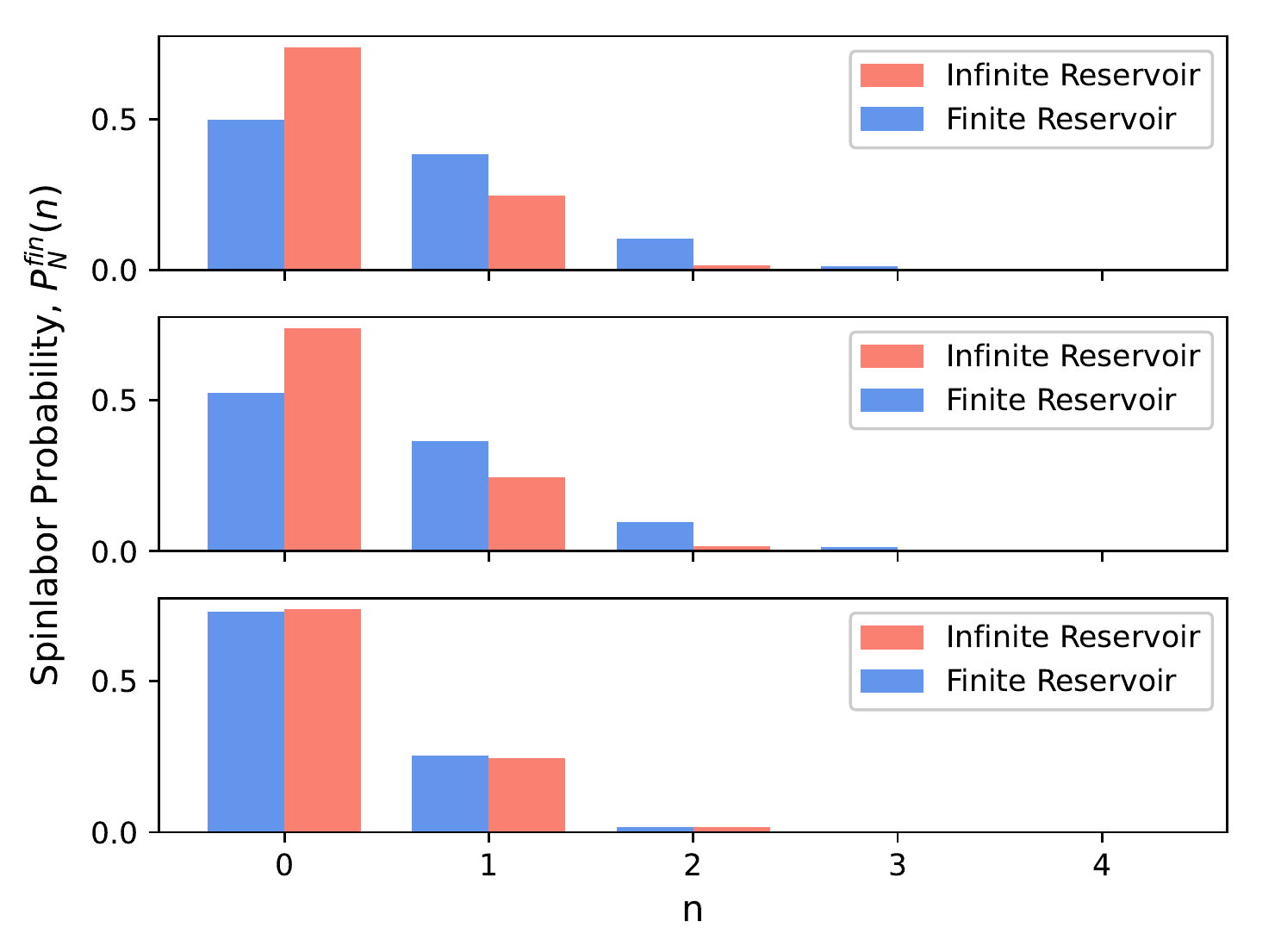}
	\caption[Spinlabor probability distribution for $\alpha_{\rm i} = 0.2$]{Spinlabor probability distribution $\P^{\text{fin}}_{N}(n)$ is given by solving \eq{eqn:recrel_fin} for $\alpha_{\rm i} = 0.2$ and $m=N$. For $\P^{\text{inf}}_{N}(n)$ we solve \eq{eqn:recrel_inf} for $\alpha= 0.2$ and $m=10000$.  The top plot has a reservoir size of $N=5$, the middle has a size of $N=10$ and the bottom plot has a reservoir size of $N=100$.}
	\label{fig:work_dist_02}
\end{figure}

\begin{figure}
	\centering
	\includegraphics[width=0.5\textwidth]{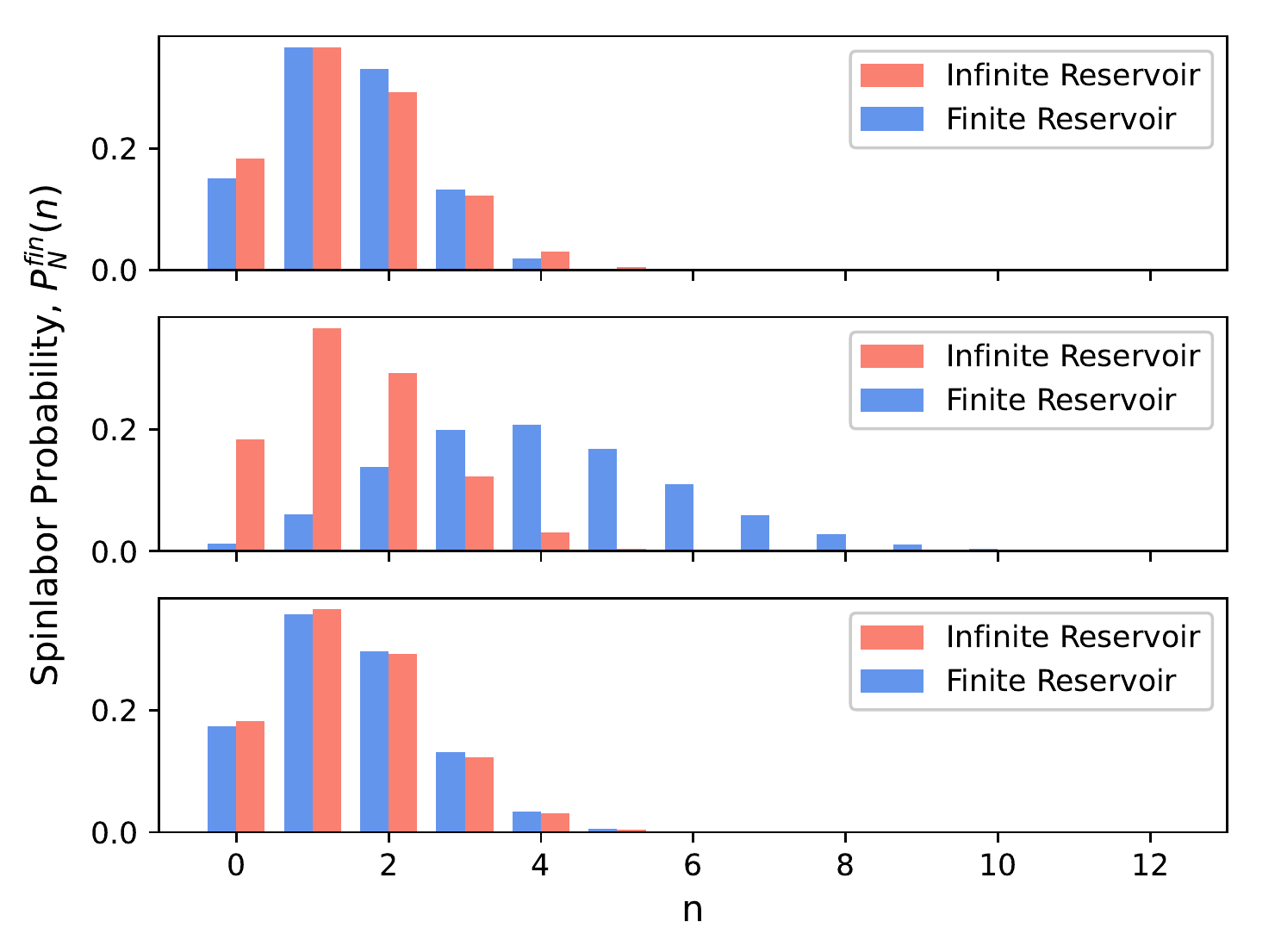}
	\caption[Spinlabor probability for $\alpha_{i} = 0.4$]{Spinlabor probability distribution for $\alpha_{\rm i} = 0.4$. The top plot has a reservoir size of $N=5$, the middle has a size of $N=100$, and the bottom plot has a reservoir size of $N=500$.}
	\label{fig:work_dist_04}
\end{figure}

\begin{figure}
	\centering
	\includegraphics[width=0.5\textwidth]{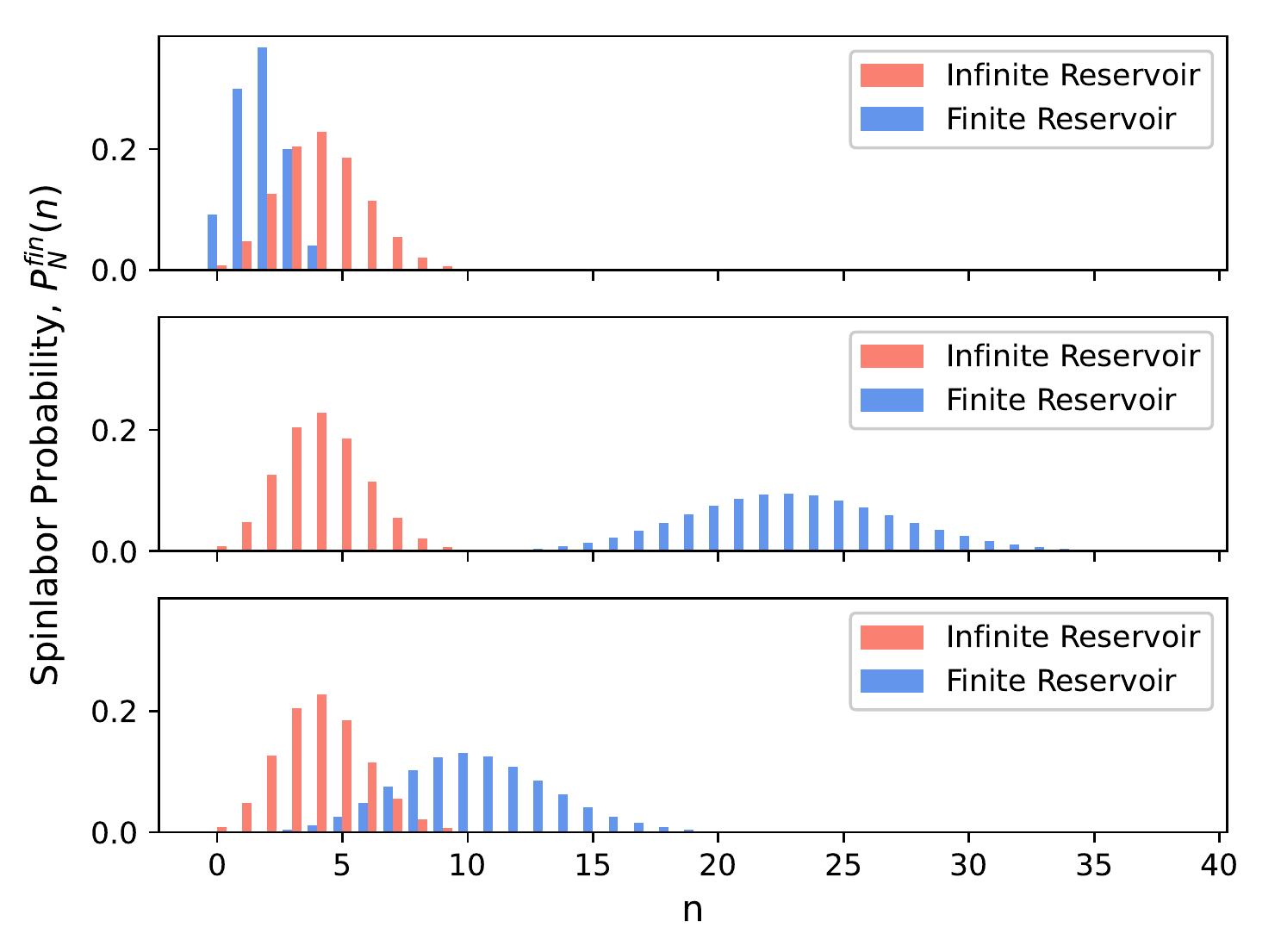}
	\caption[Spinlabor probability for $\alpha_{\rm i} = 0.46$]{Spinlabor probability distribution for $\alpha_{\rm i} = 0.46$. The top plot has a reservoir size of $N=5$, the middle has a size of $N=100$, and the bottom plot has a reservoir size of $N=1000$.}
	\label{fig:work_dist_046}
\end{figure}

Using \eqs{eqn:recrel_inf} and \eqr{eqn:recrel_fin}, we compare the spinlabor cost probability distributions for the finite and infinite reservoirs, for various finite reservoir sizes and initial spin polarisations in Figs. \ref{fig:work_dist_02}, \ref{fig:work_dist_04} and \ref{fig:work_dist_046}.
As both the spin polarisation $\alpha_{m}$ of a finite reservoir and $p_{\up, m}$ of the memory/memory-ancilla system changes during each equilibration step, we need to distinguish its values at different stages of the erasure process. In particular, we shall use subscripts ``i'' and ``f'' to label its initial and final values, respectively. In contrast, the spin polarisation of the infinite reservoir is assumed to remain unchanged throughout the erasure process, and so $\alpha$ for the reservoir remains constant.

We first examine the effect of changing the size of the finite reservoir. The three figures show that when the reservoir size is small (upper plots), the spinlabor distribution for the finite and infinite reservoir are similar in respect that they both have low costs. As the reservoir size increases (middle and bottom plots), the distributions become significantly different before tending to converge. The converging spinlabor costs in the last plot of each figure is expected because the finite reservoir tends to more-closely resemble its infinite counterpart as a thermodynamic resource as its size increases. The fact that a mid-sized reservoir has a higher spinlabor cost compared to the infinite reservoir (i.e. middle plots of Figs. \ref{fig:work_dist_04} and \ref{fig:work_dist_046}) is expected on the basis that a finite reservoir will be less efficient at erasing information. This is clearly evident from Table \ref{tab:sum_3_figs} where we see that as the reservoir size increases, $p_{\uparrow, f}$ approaches zero. Finally, the fact that the spinlabor costs are low for a small reservoir (i.e. the top plots in each figure) follows from the limited number of $N-1$ erasure cycles possible when using a reservoir with $N$ spin particles. The spinlabor cost is upper-bounded by $(N-1)\hbar$ with an average that is less than $\frac{1}{2}(N-1)\hbar$ because the probability of the cost being $\hbar$ for each cycle is less than $\frac{1}{2}$. As $N=5$ in these cases, we expect an average spinlabor cost that is bounded approximately by $2\hbar$. In comparison, the corresponding average spinlabor cost when using an infinite reservoir according to \eq{eqn:VB} and \eqr{eqn:Spinlabor_ineq} is bounded by 
\bea
  \langle\mathcal{L}_{\rm s}\rangle & \geq & \hbar \frac{\ln\left(\frac{1}{1 - \alpha}\right)}{\ln\left(\frac{1 - \alpha}{\alpha}\right)}.
    \label{eqn:Spinlabor_ineq__in_terms_alpha}
\eea
When $\alpha = \alpha_{\rm i} = 0.2$, $0.4$ and $0.46$ the spinlabor cost is bounded by $\langle\mathcal{L}_{\rm s}\rangle  \geq  0.16 \hbar$, $1.26 \hbar$, and $3.84 \hbar$, respectively.
Thus, the spinlabor costs for a very small reservoir are \emph{suppressed} compared with its infinite counterpart in the high spin-temperature regime $\alpha_{\rm i}\gtrsim 0.4$.

\begin{table}
\begin{tabular}{ |c|c|c|c| } 
\hline
$\alpha_{\rm i}$ & Reservoir Size & $p_{\uparrow,{\rm f}}$ \\
\hline
\multirow{3}{2em}{0.2} & 5 & 0.1131 \\

& 10 & 0.0340 \\ 

& 100 & 8.98e-11 \\ 
\hline
\multirow{3}{2em}{0.4} & 5 & 0.3551 \\ 

& 100 & 0.0266 \\ 

& 500 &  4.72e-06 \\ 
\hline
\multirow{3}{2em}{0.46} & 5 & 0.4414 \\ 

& 100 & 0.2216 \\ 

& 1000 & 0.0062 \\ 
\hline
\end{tabular}
\caption{Table showing the final memory probability $p_{\uparrow,{\rm f}}$ of being in the spin-up state for the different values of $\alpha_{\rm i}$ and reservoir size $N$ which correspond to the values used in Figs. \ref{fig:work_dist_02}, \ref{fig:work_dist_04} and \ref{fig:work_dist_046}.} \label{tab:sum_3_figs}
\end{table}

Next, we examine the effect of different values of the initial spin polarisation of  $\alpha_{\rm i}=0.2$, $0.4$ and $0.46$ corresponding to relatively low, mid range and relatively high spin temperatures in Figs. \ref{fig:work_dist_02}, \ref{fig:work_dist_04} and \ref{fig:work_dist_046}, respectively. We note that the rate at which the spinlabor cost for a finite reservoir converges with increasing size (corresponding to going from the top to the bottom plot of each figure) to that of an infinite reservoir slows as $\alpha_{\rm i}$ increases. There is a corresponding trend evident in Table \ref{tab:sum_3_figs} where the final probability that the memory is in the spin-up state $p_{\uparrow,\rm f}$ for the finite reservoir approaches zero slower with increasing $N$ the closer $\alpha_{\rm i}$ is to $0.5$, with the value $p_{\uparrow,\rm f} = 0$  for the infinite reservoir. Clearly, as the spin temperature increases, a finite spin reservoir needs to grow sufficiently in size in order to continue to perform like its infinite counterpart.

\begin{figure}
	\centering
	\includegraphics[width=0.5\textwidth]{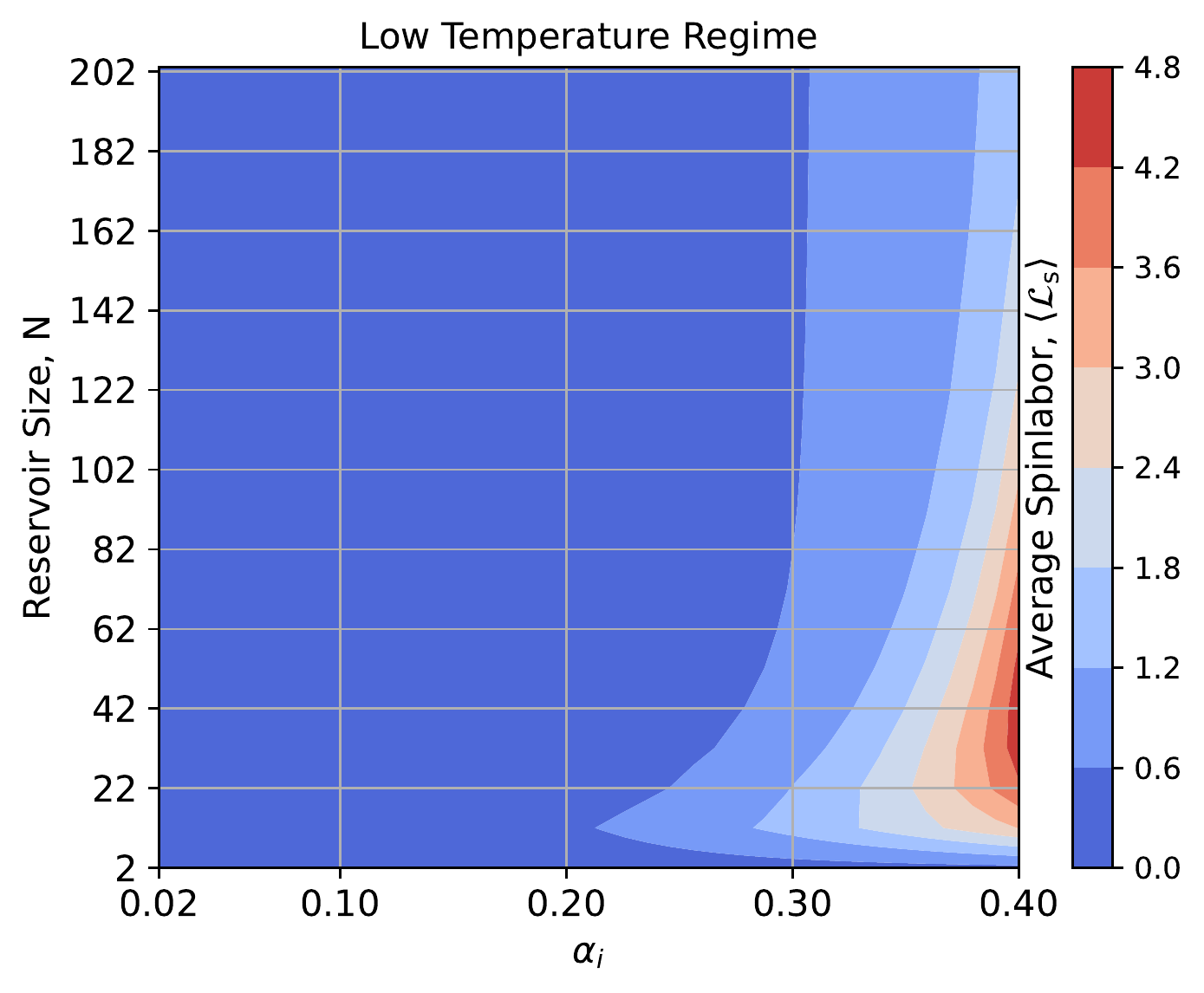}
	\caption[Average work]{Average spinlabor $\langle\mathcal{L}_{\rm s}\rangle$ defined in \eq{eqn:avg_fin} as a function of different finite reservoir sizes $N$ and initial reservoir spin polarisation $\alpha_{\rm i}$ in the low spin temperature regime of $\alpha_{\rm i}$ between $0.02$ and $0.4$. To enhance the graphical representation the values have been interpolated between the discrete values of $N$. }
	\label{fig:avg_work_low_temp}
\end{figure}

\begin{figure}
	\centering
	\includegraphics[width=0.5\textwidth]{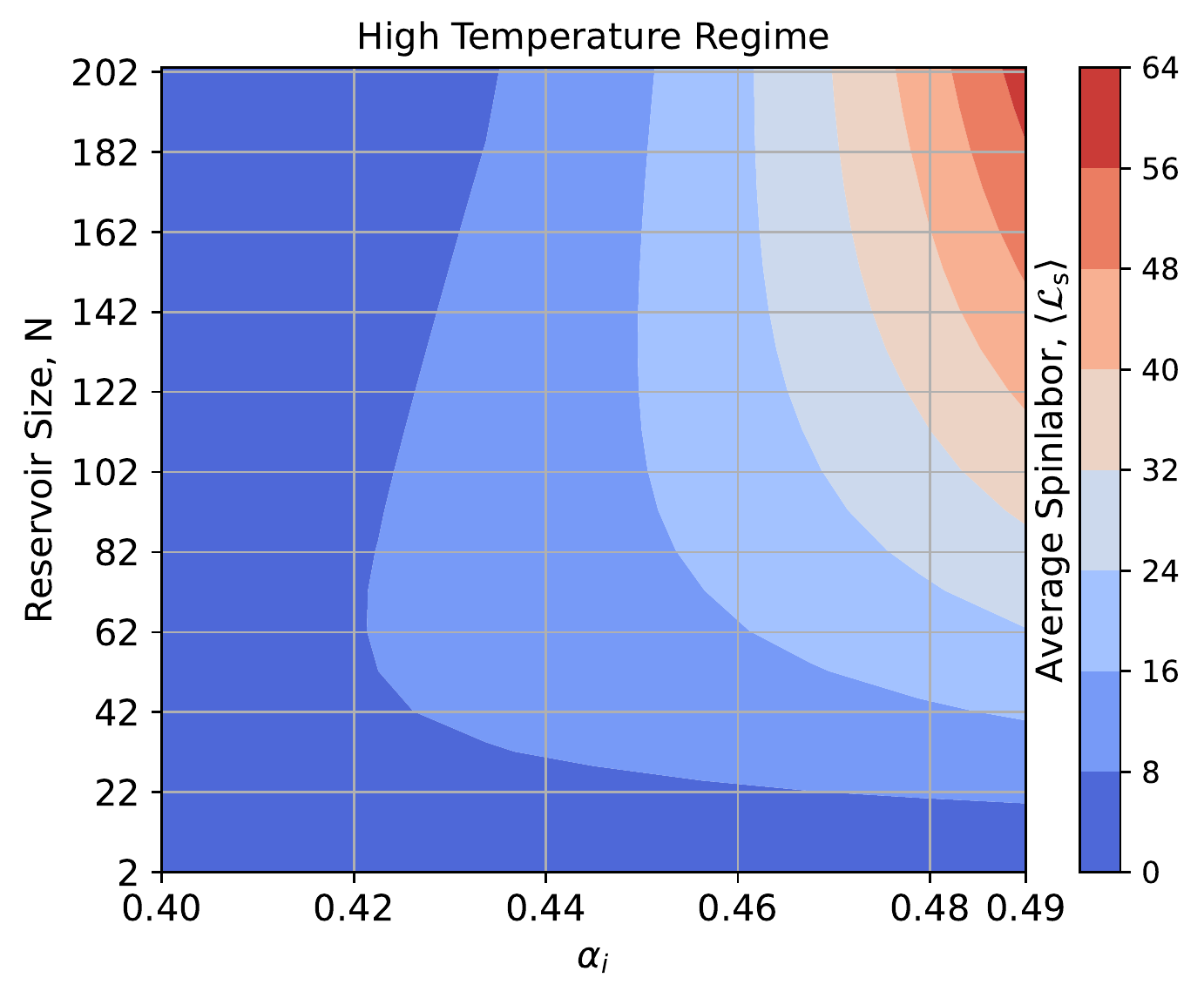}
	\caption[Average work]{Average spinlabor similar to Fig. \ref{fig:avg_work_low_temp} but in  high spin temperature regime of $\alpha_{\rm i}$ between $0.4$ and $0.49$.}
	\label{fig:avg_work_high_temp}
\end{figure}

To explore these trends in more detail, we plot the average spinlabor cost $\ip{\mathcal{L}_{\rm s}}$ for a finite reservoir given in \eq{eqn:avg_fin_other_form} as a function of reservoir size $N$ and spin polarisation $\alpha_{\rm i}$ in Figs. \ref{fig:avg_work_low_temp} and \ref{fig:avg_work_high_temp}. Fig. \ref{fig:avg_work_low_temp} represents  the initially low spin temperature regime of $\alpha_{\rm i}$ between $0.02$ to $0.4$ and Fig. \ref{fig:avg_work_high_temp} represents the initially high spin temperature regime of $\alpha_{\rm i}$ between $0.4$ to $0.49$, respectively. The first thing to note is that the suppression of the spinlabor cost for very small reservoirs is evident in both figures as the bluish regions for $N\lesssim 20$. When $\alpha_{\rm i}$ is close to $0$ in Fig. {\ref{fig:avg_work_low_temp}}, we see that the cost is low (dark blue) and only when $\alpha_{\rm i}$ becomes close to $0.4$ does the average spinlabor cost increase appreciably (light blue). Near $\alpha_{\rm i} = 0.4$, we find that the average spinlabor cost is significantly higher (pink to red) for mid range values of $N=20$ to $100$. 

\begin{figure}
	\centering
	\includegraphics[width=0.5\textwidth]{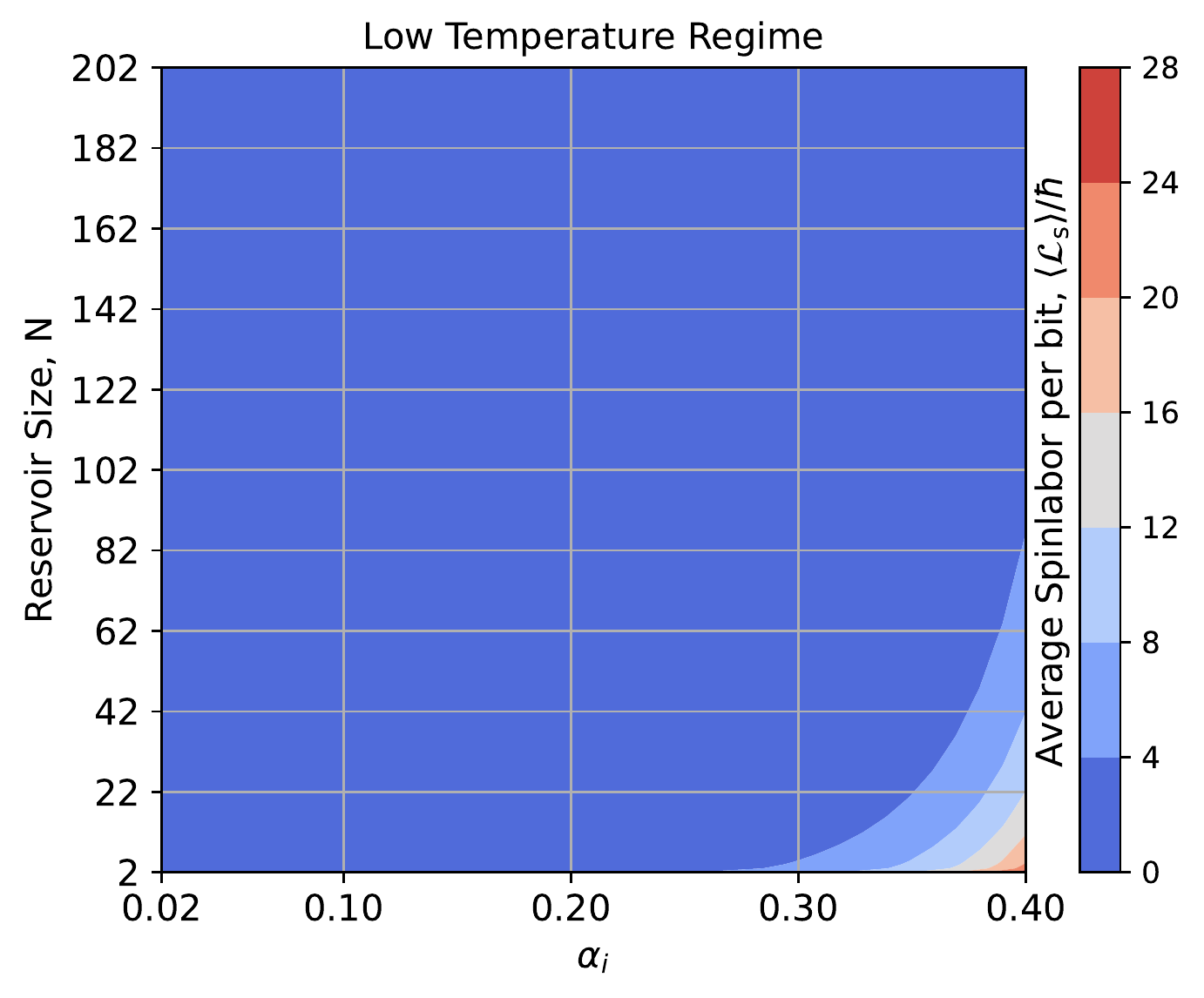}
	\caption[Average work per bit]{Average spinlabor per bit erased defined by \eq{eqn:spinlabor_per_bit} as a function of different finite reservoir sizes $N$ and initial reservoir spin polarisation $\alpha_{\rm i}$ for the low spin temperature regime of $\alpha_{\rm i}$ between $0.02$ and $0.4$. To enhance the graphical representation the values have been interpolated between the discrete values of $N$. }
	\label{fig:avg_work_per_bit_low_temp}
\end{figure}

\begin{figure}
	\centering
	\includegraphics[width=0.5\textwidth]{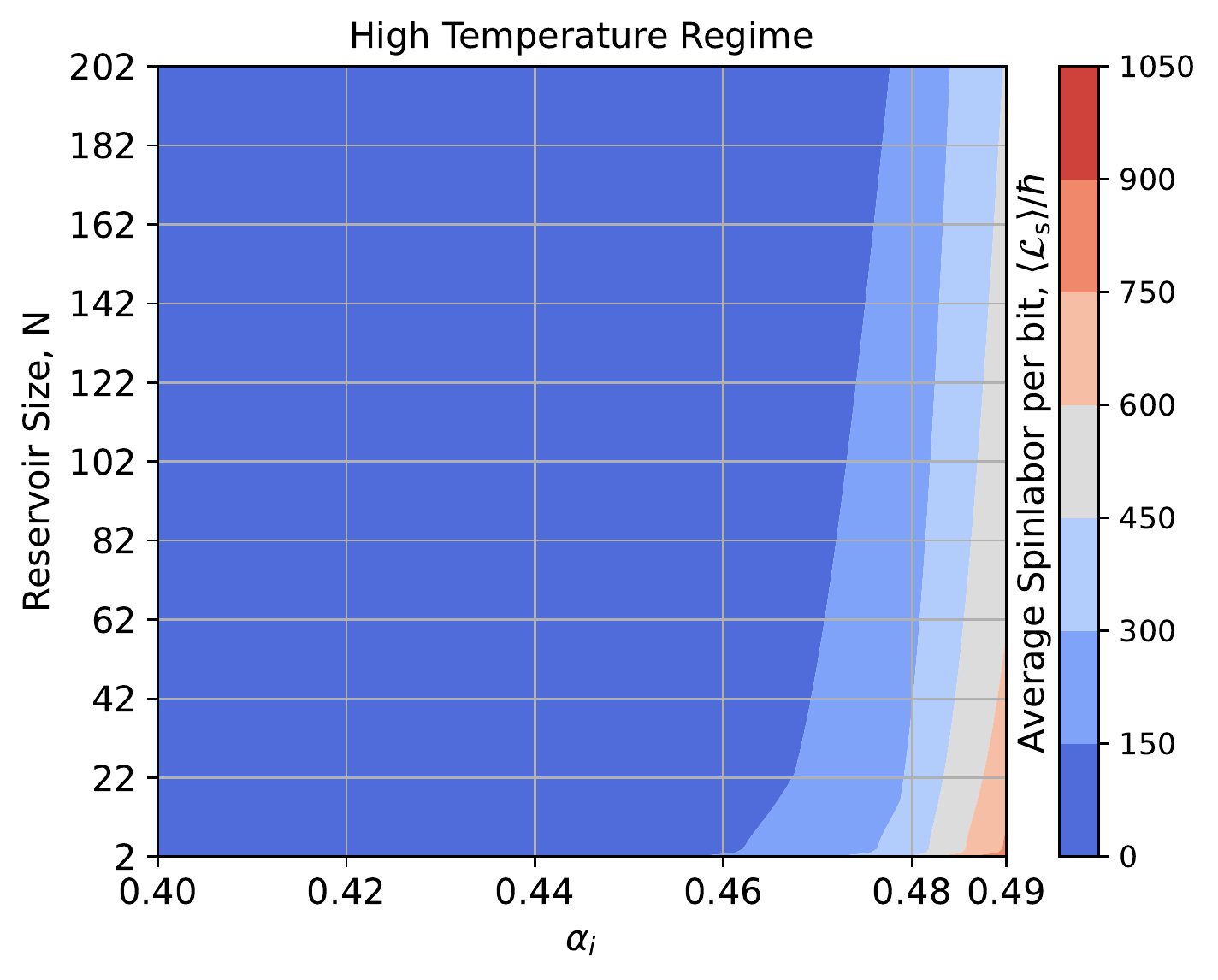}
	\caption[Average work per bit]{Average spinlabor per bit erased similar to Fig. \ref{fig:avg_work_high_temp} but for the high spin temperature regime of $\alpha_{\rm i}$ between $0.4$ and $0.49$.}
	\label{fig:avg_work_per_bit_high_temp}
\end{figure}

Evidently, this relatively large cost is due to the reservoir size being insufficient for effective erasure to occur for relatively small reservoirs and moderate spin temperatures, verifying the conclusions made from the previous figures. Fig. \ref{fig:avg_work_high_temp} shows similar trends at larger values of $N$. The entropy stored in a memory can be completely erased using an infinite reservoir whereas, the same degree of erasure is forbidden for a finite reservoir, consistent with the 3rd law of thermodynamics \cite{Scharlau2018quantumhornslemma}.
The spinlabor cost examined above are irrespective of the degree to which the erasure is performed.  A more appropriate measure of the cost of erasure is given by the  average spinlabor \emph{per bit of entropy erased}, $\langle{\mathcal{L}\mkern-7.5mu/}_{\rm s}\rangle$, which is defined as
\bea
\langle{\mathcal{L}\mkern-7.5mu/}_{\rm s}\rangle = \langle\mathcal{L}_{\rm s}\rangle\frac{ \ln 2}{\ln 2 + p_{\downarrow,\rm f} \ln p_{\downarrow,\rm f} + p_{\uparrow,\rm f} \ln p_{\uparrow,\rm f}}
\label{eqn:spinlabor_per_bit}
\eea
for a memory that initially contains 1 bit of entropy (i.e. $\ln2$ nats) where $p_{\uparrow,\rm f}$ and $p_{\downarrow,\rm f}$ are the final values of the probabilities that the memory-ancilla system is in the spin-up and -down states, respectively, at the end of the erasure process.  The inverse of the last factor on the right side is the change in the entropy of the memory-ancilla system, expressed in bits. Figs. \ref{fig:avg_work_per_bit_low_temp} and \ref{fig:avg_work_per_bit_high_temp} plot the average spinlabor per bit erased, $\langle{\mathcal{L}\mkern-7.5mu/}_{\rm s}\rangle$, for an initially low spin temperature regime of $\alpha_{\rm i}$ between $0.02$ to $0.4$ and initial high spin temperature regime of $\alpha_{\rm i}$ between $0.4$ to $0.49$ respectively. Both plots show that the average spinlabor cost per bit reduces with increasing reservoir size, indicating that the erasure process becomes more efficient per bit erased as the size of the reservoir grows, as expected. We note that this indicates that in \eq{eqn:avg_fin} the rate at which $p_{\uparrow,\rm f}$ decreases is faster than the rate at which $N$ increases and is why the average spinlabor cost approaches a particular value as $N \rightarrow \infty$. In addition, we find that as the spin temperature \eq{eqn:gamma} of the reservoir increases (i.e. as $\alpha_{\rm i}$ increases and approaches $0.5$) more spinlabor is required to erase the same amount of information.

\subsection{Matching Infinite and Finite Reservoir}
While the \emph{average} values are important when comparing costs, they are not the only things to consider.  To place the comparison of the finite and infinite reservoirs on an equal footing we need to ensure that the statistics of the costs are comparable as well. We utilize the Jensen-Shannon Divergence (JSD) \cite{Lin1991} for this. The JSD is a symmetric measure of how close two probability distributions $P=\{P(x):x\in X\}$ and $Q=\{Q(x):x\in X\}$ are to each other, and is defined as
\bea
\mbox{JSD}(P||Q) = \frac{\mbox{D}(P||M) + \mbox{D}(Q||M) }{2}
\eea
where $M = \frac{1}{2}(P+Q)$ and $\mbox{D}(P||M)$ or $\mbox{D}(Q||M)$ is the Kullback-Leibler divergence \cite{Kullback1951} expressed as
\bea   \label{eqn:Kullback-Leibler divergence}
\mbox{D}(P||M) = \sum_{x \in X} P(x) \log \frac{P(x)}{M(x)}.
\eea
$\mbox{JSD}(P||Q)$ is bounded such that 
\bea
0 \leq \mbox{JSD}(P||Q) \leq \ln 2,
\eea
where the closer $\mbox{JSD}(P||Q)$ is to $0$ the closer the distributions are to each other. A value of $\mbox{JSD}(P||Q) = 0$ indicates that the probability distribution are exactly the same. 

In our case, we wish to compare the spinlabor costs distributions for the finite  and infinite reservoirs, $\mathcal{P}^{\rm fin}_N$ in \eq{eqn:recrel_fin} and $\mathcal{P}^{\rm inf}_\infty$ in \eq{eqn:recrel_inf}, respectively.  Given that $\mathcal{P}^{\rm fin}_N \to \mathcal{P}^{\rm inf}_\infty$ as $N\to\infty$, it follows that $\mbox{JSD}(\mathcal{P}^{\rm fin}_N||\mathcal{P}^{\rm inf}_\infty)\to 0$  as $N\to\infty$, i.e. the spinlabor cost distributions are \emph{indistinguishable} in the large $N$ limit.  A more practical issue is to determine the smallest size $N$ of a finite reservoir that has costs and performance that are comparable to that of the equivalent infinite reservoir. 
Let $\Delta{\rm JSD}$ be the upper bound of  $\mbox{JSD}(\mathcal{P}^{\rm fin}_N||\mathcal{P}^{\rm inf}_\infty)$ that we can tolerate.
The smallest value of $N$ that gives spinlabor statistics that differ tolerably from those of the infinite reservoir is given by  
\begin{align}  \label{eqn:N-min}
    N^{\rm min}_{\Delta{\rm JSD}} =  \min_{N} \{N:\mbox{JSD}(\mathcal{P}^{\rm fin}_N||\mathcal{P}^{\rm inf}_\infty) < \Delta{\rm JSD}\}.
\end{align}
In the remainder of this work, we choose $\Delta\mbox{JSD}=0.005$ nats, i.e. our tolerance criterion is that the spinlabor cost distributions differ by less than $0.005$ nats as measured by the JSD. Note that we want to calculate the minimum reservoir size out of a set of reservoir sizes that satisfy the $\Delta\mbox{JSD}$ tolerance criterion and that all subsequent reservoir sizes that are greater than the calculated reservoir size meet this condition as well. Our chosen value of $0.005$ nats provides us with the first reservoir size calculated by $\Delta\mbox{JSD}$ tolerance criterion to be the one that satisfies the above conditions.

\begin{figure}
	\centering
	\includegraphics[width=0.5\textwidth]{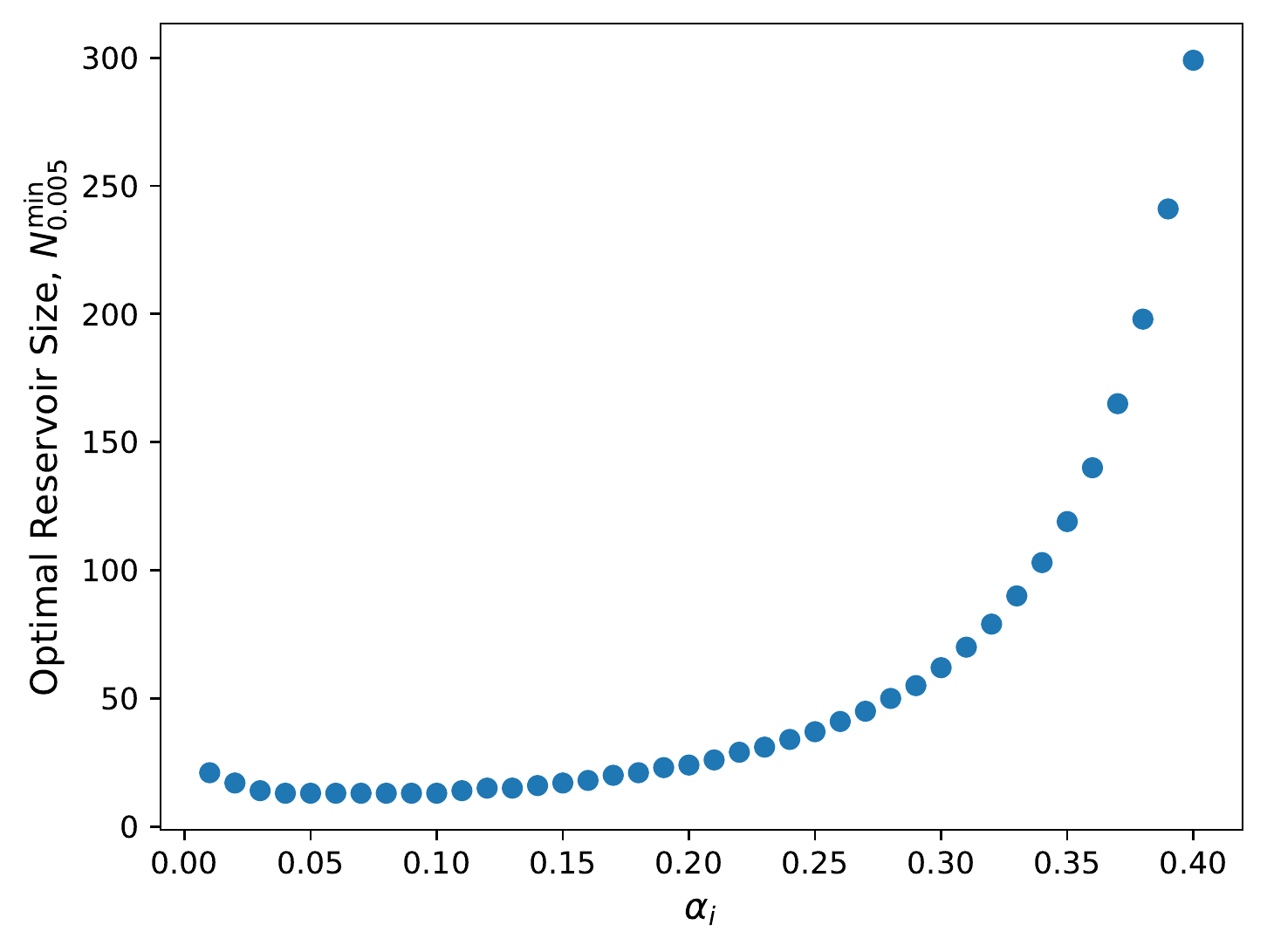}
	\caption[Similarity of JSD $< 0.005$ comparing finite and infinite reservoir spinlabor probability]{The minimum reservoir size $N^{\rm min}_{\Delta{\rm JSD}}$ defined in \eq{eqn:N-min} as a function of $\alpha_{\rm i}$ which ranges from 0.01 to 0.40 in intervals of 0.01 for JSD tolerance level $\Delta\mbox{JSD}=0.005$.}
	\label{fig:similarity_between_work_prob}
\end{figure}

\begin{figure}
	\centering
	\includegraphics[width=0.5\textwidth]{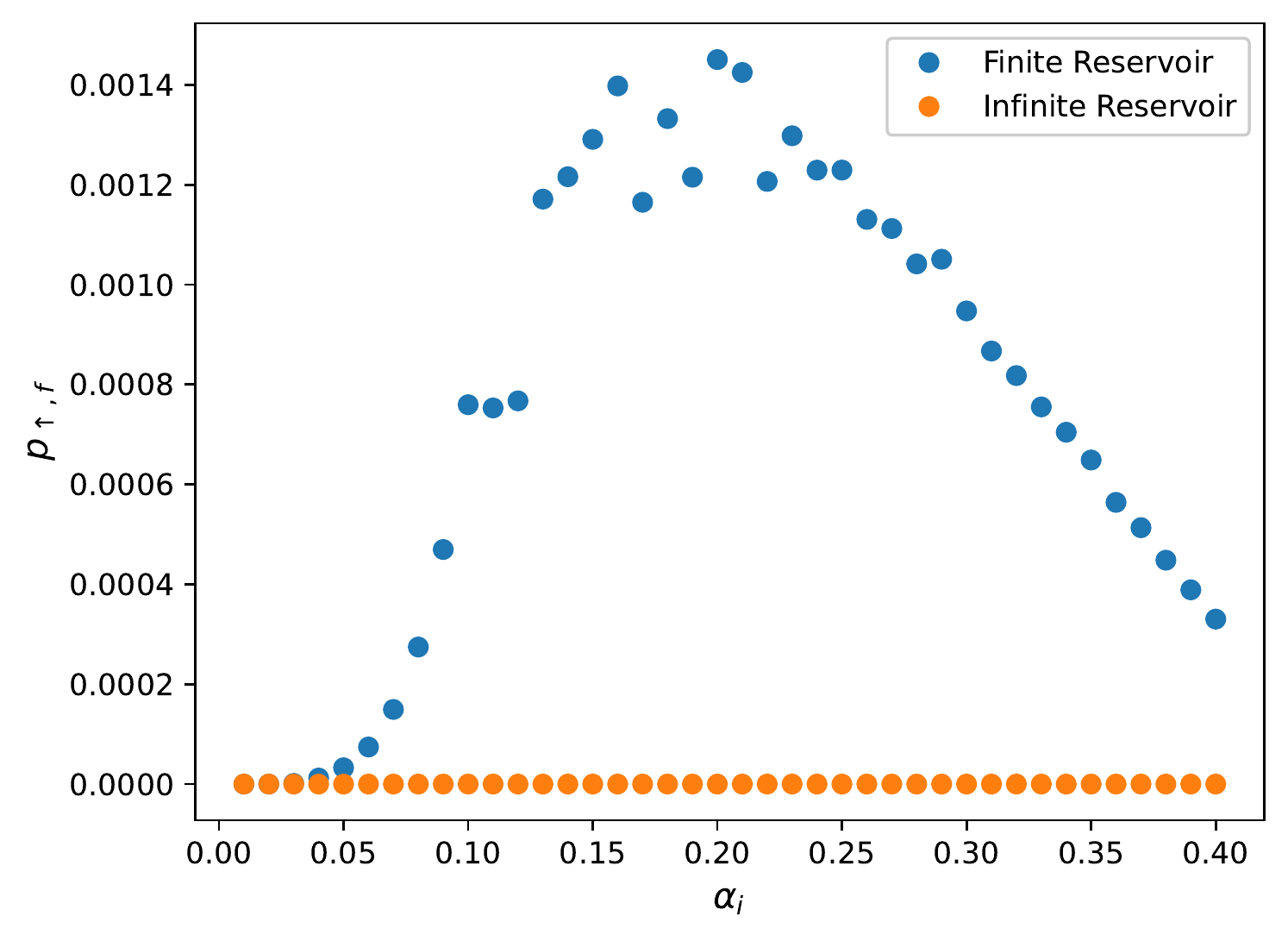}
	\caption[Comparing erasure]{Probability $p_{\uparrow,\rm f}$ of the memory being in the spin-up state at the end of the erasure process, with values of $\alpha_{\rm i}$ and the corresponding reservoir size $N^{\rm min}_{\Delta{\rm JSD}}$ the same as in Fig. \ref{fig:similarity_between_work_prob}.}
	\label{fig:similarity_between_erasure}
\end{figure}

We have performed a numerical analysis of $N^{\rm min}_{\Delta{\rm JSD}}$ for a set of values of $\alpha_{\rm i}$. To compare like to like we have limited the number of erasure cycles for the infinite reservoir to be the same as for the finite reservoir and the results are plotted in Fig. \ref{fig:similarity_between_work_prob}. The figure shows $N^{\rm min}_{\Delta{\rm JSD}}$ increases monotonically with $\alpha_{\rm i}$ except for values of $\alpha_{\rm i}$ that are close to zero as seen in Fig. \ref{fig:similarity_between_work_prob} for $\alpha_{\rm i}\le 0.03$. 
We note that in the ultra cold limit $\alpha_{\rm i}\to 0$, an infinite reservoir would perfectly erase the memory after just the first equilibration step, and so the spinlabor cost would be zero and have the extreme distribution $\mathcal{P}^{\rm inf}_\infty(n)=\delta_{n,0}$. The spinlabor distribution in the case of a finite reservoir in the same ultra cold limit would approach the extremely narrow distribution $\delta_{n,0}$ only if its size $N$ is extremely large.  Thus, it should not be unexpected that $N^{\rm min}_{\Delta{\rm JSD}}$ increases as $\alpha_{\rm i}\to 0$ in Fig. \ref{fig:similarity_between_work_prob}.  What is perhaps surprising is that a reservoir size of 24 or smaller can produce similar spinlabor distributions to that of an infinite reservoir for the moderately cold regime $0\lesssim \alpha_{\rm i}\le 0.2$ (excluding the ultra cold regime $\alpha_{\rm i}\approx 0$) as seen in Table \ref{tab:table_data} of appendix \ref{sec:Table}. Here, Table \ref{tab:table_data} presents all the values of $\alpha_{\rm i}$, minimum reservoir size $N^{\rm min}_{\Delta{\rm JSD}}$, and $p_{\uparrow,\rm f}$ for the finite reservoir that were used in plotting Figs. \ref{fig:similarity_between_work_prob} and \ref{fig:similarity_between_erasure}. 

However, even though the spinlabor cost probability distributions of the finite and infinite reservoir are similar, the degree of erasure may not be. In Fig. \ref{fig:similarity_between_erasure} we plot the probability that the memory-ancilla system is in the spin-up state at the end of the erasure process, $p_{\uparrow,\rm f}$, for both the finite and infinite reservoirs. As expected for the infinite reservoir $p_{\uparrow,\rm f}=0$ which indicates perfect erasure. Perfect erasure for finite reservoirs is not possible as a result of the third law of thermodynamics \cite{Scharlau2018quantumhornslemma}.  The fluctuating values of $p_{\uparrow,\rm f}$ for the finite reservoir are associated with the discreteness of the reservoir size and is more pronounced for $\alpha_{\rm i}\lesssim 0.2$  which, according to Fig. \ref{fig:similarity_between_work_prob} and Table \ref{tab:table_data} of appendix \ref{sec:Table}, corresponds to relatively small reservoirs $N^{\rm min}_{\Delta{\rm JSD}}\lesssim 24$.

\section{The Cost of  Resetting the Ancilla \label{sec:ancilla retrival}}
When the erasure protocol is implemented using an infinite reservoir, not only is the erasure perfect in the sense that $p_{\uparrow,\rm f}=0$ and $p_{\downarrow,\rm f}=1$, but all the ancilla spins are simultaneously returned to their spin-down state $\ket{\dn}\breaka{\dn}$ (logical $0$) as well. This means that the same ancilla spins can be \emph{reused} in repeated implementations of the infinite-reservoir erasure protocol. However, this is not the case for an implementation using a finite reservoir for which $p_{\uparrow,\rm f}\ne 0$ and $p_{\downarrow,\rm f}\ne 1$, on principle \cite{Scharlau2018quantumhornslemma}.  As all ancilla spins share the same state as the memory spin, each one has $p_{\downarrow,\rm f}\ne 1$. As such, the ancilla spins cannot be reused in a further implementation of the protocol and so, without any further processing, they would represent a consumable resource that needs to be accounted for in addition to the spinlabor cost of the CNOT operations. We now show how all the ancilla spins can, in fact, be \emph{reset} to their initial spin-down state at the cost of expending an additional amount of spinlabor.

We use the fact that at the end of the erasure process, the memory spin and each of the ancilla spins are perfectly correlated due to the action of $m=N-1$ previous CNOT steps. Again, we note that for a finite reservoir we perform the same number of erasure cycles as the reservoir size will permit and hence $m=N-1$. If all spins are in the spin-down state $\ket{\dn}\breaka{\dn}$ (logical $0$) we need do nothing, and if all spins are in the spin-up state $\ket{\up}\breaka{\up}$ (logical $1$), we need to flip the state of the ancilla spins.  Thus, the operation we need entails applying $m$ CNOT operations, each one using the memory spin as the control bit and one of the ancilla spins as the target bit.  The operation will leave all the ancilla spins in the logical 0 state $\ket{\dn}\breaka{\dn}$ without changing the state of the memory spin. The average spinlabor cost of this operation is, hence,  
\bea
\langle\Delta\mathcal{L}_{\rm s}\rangle = \hbar(N-1) p_{\uparrow,N-1}  \label{eqn:get_ancilla},
\eea
With the total average spinlabor cost of erasure and resetting the ancillas to be a sum of \eqs{eqn:avg_fin} and \eqr{eqn:get_ancilla}
\bea
\langle\mathcal{L}_{\rm s}\rangle = \sum_{m=1}^{N-1} \hbar p_{\up, m-1} + \hbar(N-1) p_{\uparrow,N-1} \label{eqn:tot_avg_spin}.
\eea
This additional cost in spinlabor will change the spinlabor distribution as well. Recognizing that there is a probability of $p_{\dn,N-1}$ that the additional cost is zero, and a probability of $p_{\up,N-1}$ that it is $(N-1)\hbar$, we find that the total spinlabor distribution, after the resetting operation, is calculated as follows
\bea
    \P^{\text{fin}}_{N}(n)&=&\P^{\text{fin}}_{N-1}(n)p_{\downarrow, N-1}\nonumber\\&&\quad +\P^{\text{fin}}_{N-1}(n+N-1)p_{\up, N-1} \label{eqn:reset_spin_dist}
\eea
where $\P^{\text{fin}}_{N-1}(\cdot)$ is the spinlabor distribution before the reset operation (i.e. it is the result of iterating the recurrence relation \eq{eqn:recrel_fin} $(N-1)$ times).

\begin{figure}
	\centering
	\includegraphics[width=0.5\textwidth]{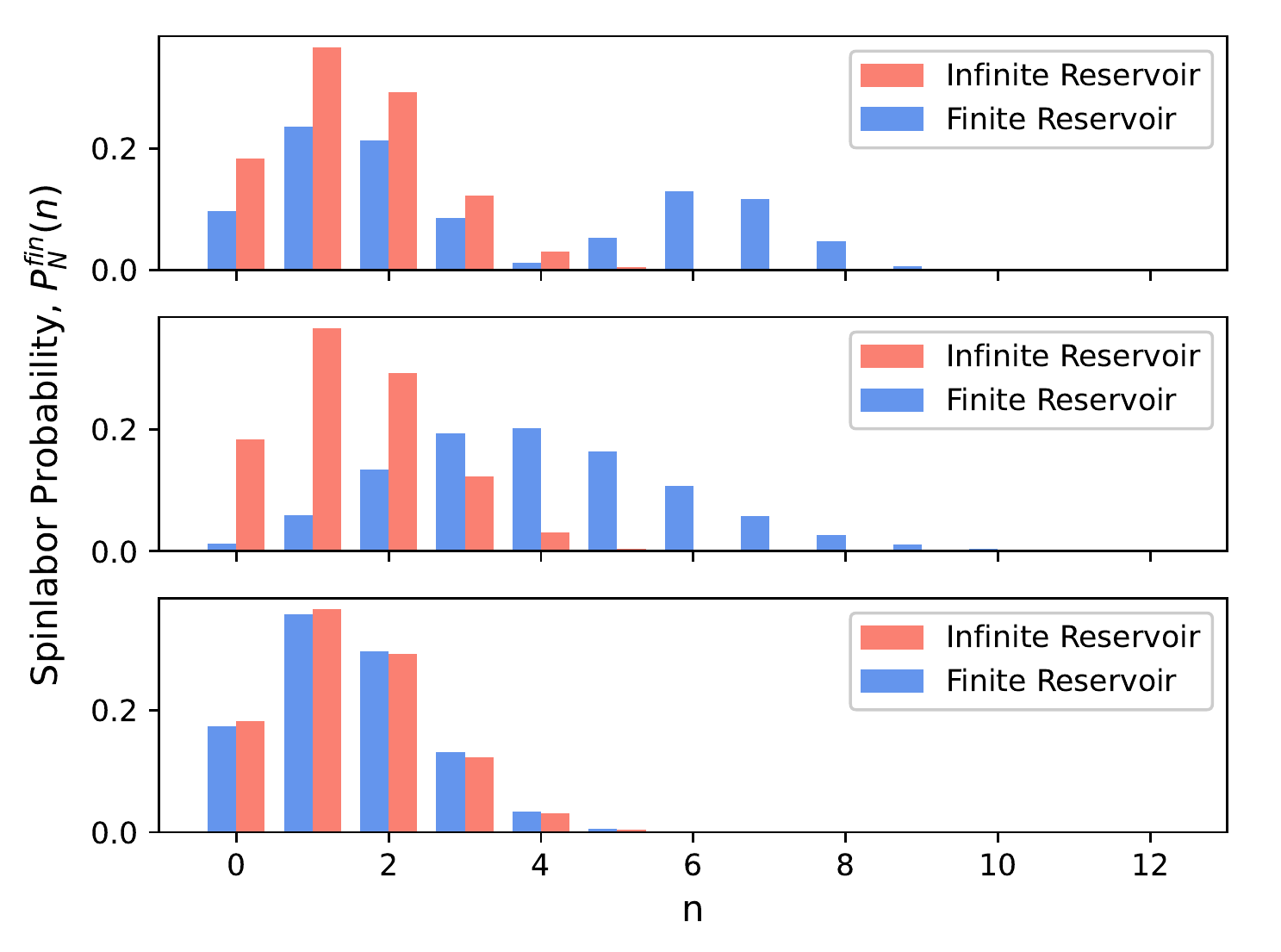}
	\caption{Spinlabor probability distribution for $\alpha_{\rm i} = 0.4$. The top plot has a reservoir size of $N=5$, the middle has a size of $N=100$, and the bottom plot has a reservoir size of $N=500$.}
	\label{fig:work_dist_plus_retrival_04}
\end{figure}

\begin{figure}
	\centering
	\includegraphics[width=0.5\textwidth]{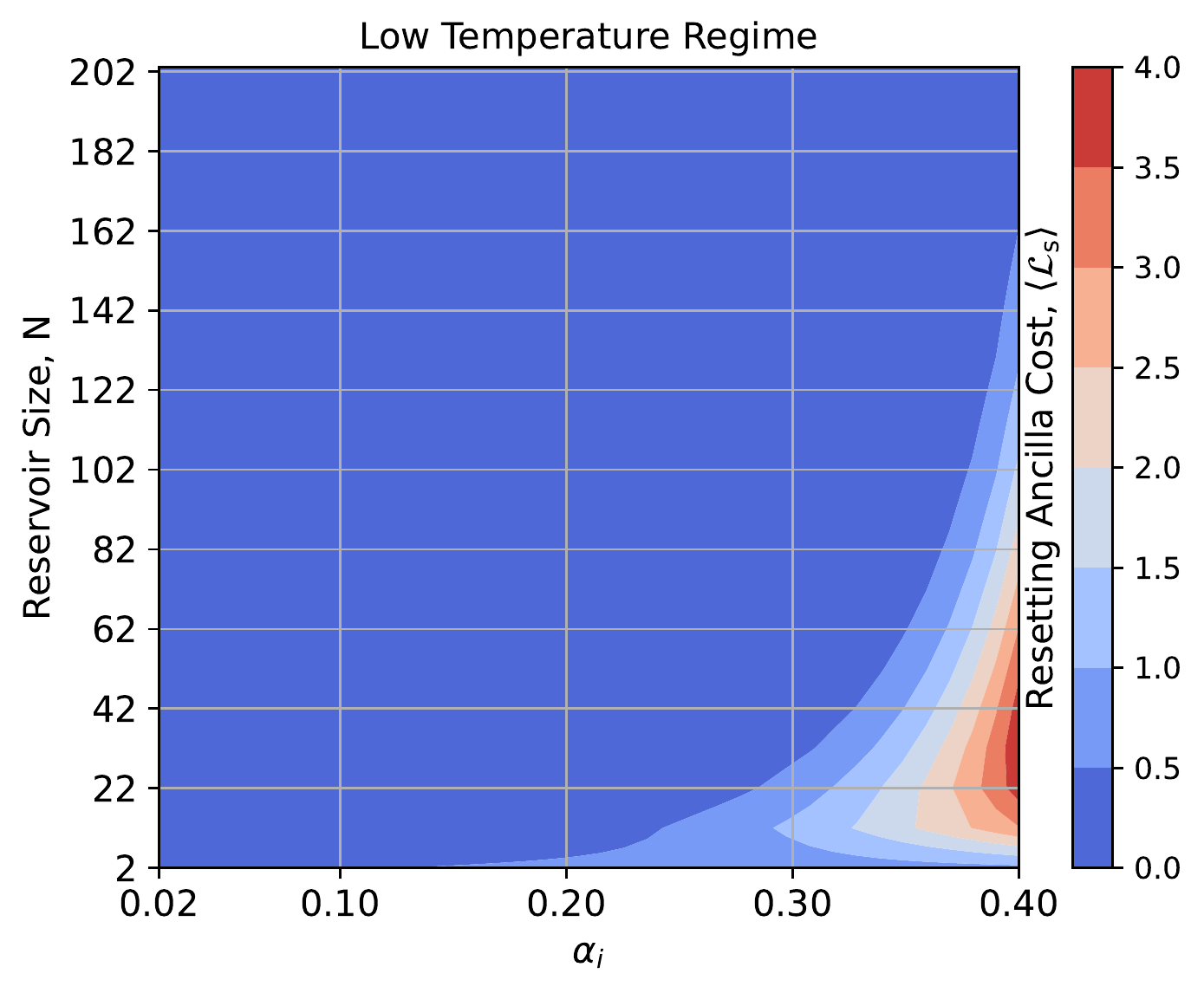}
	\caption{Average spinlabor of resetting the ancilla $\langle\mathcal{L}_{\rm s}\rangle$ defined in \eq{eqn:get_ancilla} as a function of different finite reservoir sizes $N$ and initial reservoir spin polarisation $\alpha_{\rm i}$ in the low spin temperature regime of $\alpha_{\rm i}$ between $0.02$ and $0.4$. To enhance the graphical representation the values have been interpolated between the discrete values of $N$.  }
	\label{fig:avg_work_reset_low_temp}
\end{figure}

\begin{figure}
	\centering
	\includegraphics[width=0.5\textwidth]{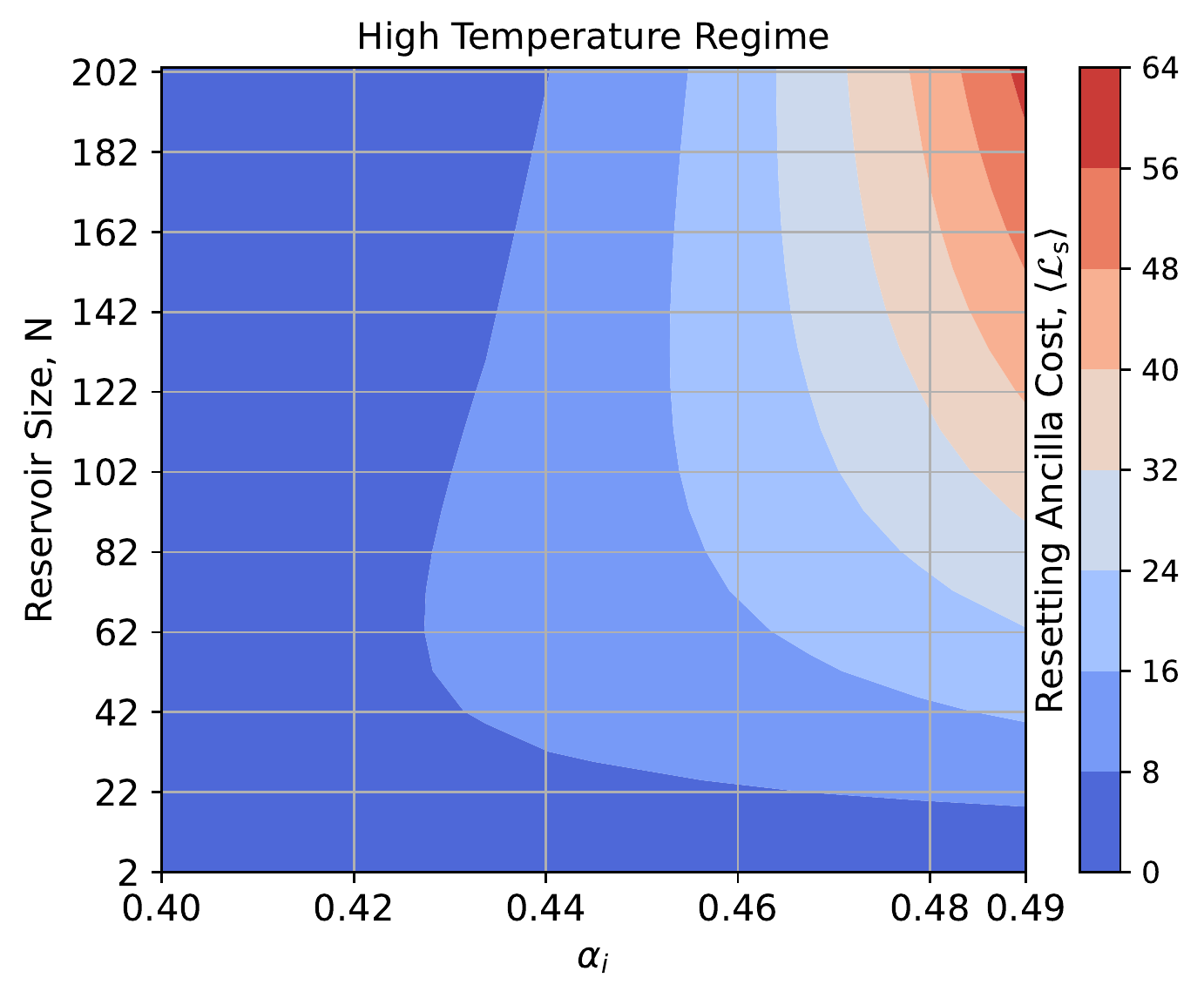}
	\caption{Average spinlabor similar to Fig. \ref{fig:avg_work_reset_low_temp} but in  high temperature regime of $\alpha_{\rm i}$ between $0.4$ and $0.49$.}
	\label{fig:avg_work_reset_high_temp}
\end{figure}

Fig. \ref{fig:work_dist_plus_retrival_04} plots the distribution for the total spinlabor cost (including the ancilla resetting cost) $\P^{\text{fin}}_{N}(n)$ as a function of $n$  for $\alpha_{\rm i} = 0.4$ for different reservoir sizes $N$.  Interestingly, it is bimodal in the $N=5$ case (top plot).  We also see that as the reservoir size increases the distribution converges to the infinite reservoir distribution, as expected. An interesting feature to investigate is the average of this additional cost. In Fig. \ref{fig:avg_work_reset_low_temp} and Fig. \ref{fig:avg_work_reset_high_temp} we plot the average spinlabor cost of resetting the ancilla. Although the costs are less than the average spinlabor cost before the resetting operation, theses plots appear to be similar to Fig. \ref{fig:avg_work_low_temp} and Fig. \ref{fig:avg_work_high_temp}. Re-expressing the right side of \eq{eqn:avg_fin} in terms of the difference $\Delta_{m}\equiv p_{\up, m} - p_{\up, N-1}$ between $p_{\up, m}$ and $p_{\up, N-1}$,   we find
\bea
\langle\mathcal{L}_{\rm s}\rangle &=& \sum_{m=1}^{N-1} \hbar p_{\up, m-1}\\ \nonumber
&=& \hbar p_{\up, N-1} +  \hbar p_{\up, N-1} + \hbar\Delta_{N-2} + . \; . \; \\ \nonumber
&& . \; . \; + \hbar p_{\up, N-1} + \hbar \Delta_{1} + \\ \nonumber
&&  + \hbar p_{\up, N-1} + \hbar\Delta_{0}  \\ \nonumber
&=&   \hbar (N-1) p_{\up, N-1} + \sum_{m=1}^{N-1} \hbar \Delta_{m-1}
\label{eqn:tot_avg_spin rearranged},
\eea
which is greater than the average reset cost in \eq{eqn:get_ancilla} because $\Delta_m>0$ for all $m$. Unfortunately, the consequence of this is that if $\sum_{m=1}^{N-1} \hbar \Delta_{m-1}$ is close to zero the total spinlabor cost would almost be doubled. To avoid this worse-case scenario we need to minimize $p_{\uparrow,N}$. Luckily this is the goal of our erasure process.  In fact, minimizing $p_{\up,N-1}$ minimizes the cost of resetting the state of the ancilla in any case, and this can be achieved by choosing either large reservoir sizes or low spin temperatures (i.e. small $\alpha_{\rm i}$ values) or both.  

\section{Reusing Reservoir \label{sec:reusing reservoir}}

\begin{figure}
	\centering
	\includegraphics[width=0.5\textwidth]{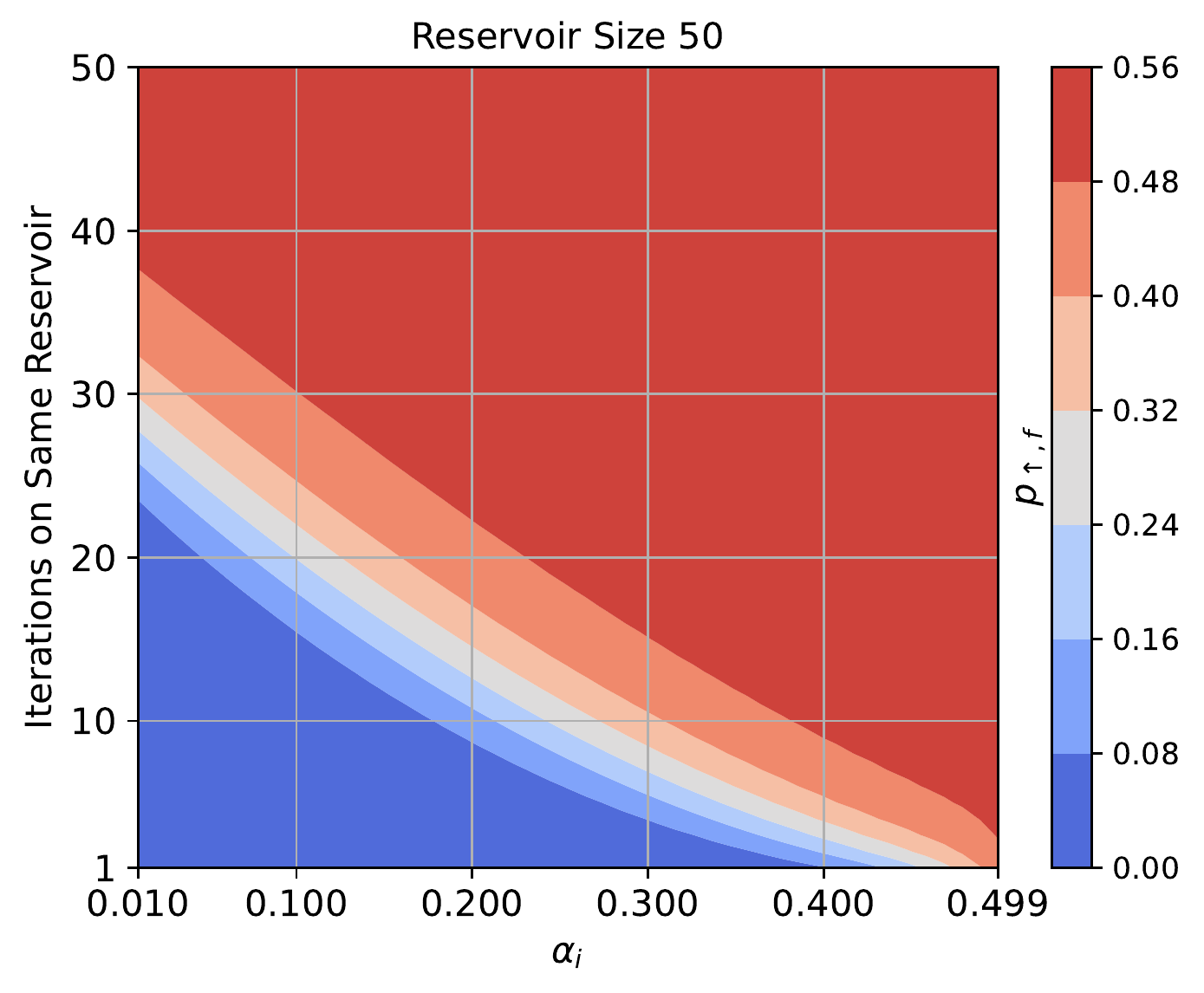}
	\caption[prob_up_reuse_reservoir]{Final probability $p_{\uparrow,\rm f}$ of the memory being in the spin-up state. Here the reservoir's attributes are recorded and reused for the next iteration with 50 iterations in total completed for a reservoir size of 50.}
	\label{fig:prob_up_reuse_reservoir}
\end{figure}

Finally, we analyze the effect of reusing the same reservoir for the erasure of additional memory spins. Regardless of how many times it is reused, an infinite reservoir retains a fixed spin temperature, in principle, and so it performs consistent erasure each time. In contrast, the spin temperature of a finite reservoir \emph{necessarily rises} as it absorbs the entropy of the memory and associated spintherm during each erasure process, where spintherm is the spin equivalent of heat \cite{Croucher2017,Croucher2021}.  As illustrated by the plots in Figs. \ref{fig:avg_work_low_temp} and \ref{fig:avg_work_high_temp}, the spinlabor cost increases with the spin temperature of the reservoir, and so the cost associated with a finite reservoir will necessarily increase each time it is reused. The final state of the reservoir after the protocol is completed (i.e. for $m=N-1$) is given by
\bea
P_{\uparrow}(n) = \sum_{M=0}^{1} P_{N-1}(n,M), \label{eqn:reservoir_prob}
\eea
where $P_{m}(n,M)$ is the joint probability distribution for the reservoir and memory defined in \eq{eqn:sum_degen}. The process of reusing the reservoir to erase the information in an additional memory is modeled by (1) replacing the initial distribution $P_{\uparrow}(n)$, that is derived in \eq{eqn:P_up(n)} using \eq{eqn:probr}, with the distribution calculated in \eq{eqn:reservoir_prob} together with the initial probabilities $p_{\up}=p_{\dn}=\frac{1}{2}$ for the new memory particle in the initial conditions \eqs{eqn:res_mem_prob_down m=0} and \eqr{eqn:res_mem_prob_up m=0}, and then (2) using the recurrence relations \eqs{eqn:res_mem_prob_down} and \eqr{eqn:res_mem_prob_up} as before to determine new values for $P_m(n,M)$ for $m=1,2,\dotsc,N-1$. The result is an additional iteration of the erasure protocol that reuses the same reservoir to erase a fresh memory particle. Fig. \ref{fig:prob_up_reuse_reservoir} plots the final probability of the memory being in the spin up state $p_{\uparrow,\rm f}$ for 50 iterations of using a single reservoir in this way and shows that when the reservoir is reused the reservoir becomes less efficient in erasing the information of the memory. 
In fact, this is direct evidence of the deleterious effect associated with the spintherm cost of erasure that we have  discussed in a previous paper \cite{Croucher2021}: 
the spintherm cost is the amount of spintherm transferred from the
memory-ancilla system to the spin reservoir. It is regarded as
a cost because it reduces the spin polarization of the reservoir
and, thus, the ability of a finite reservoir to act as an entropy sink for future erasure processes. Fig. \ref{fig:prob_up_reuse_reservoir} clearly illustrates this loss in erasure ability.

\section{Conclusion \label{sec:conclusion}}
This work provides a more practical analysis on the erasure protocol in Ref \cite{Vaccaro2011, Barnett2013, Croucher2017, Croucher2021}. We presented an analysis on information erasure for a finite spin reservoir. The erasure protocol used spin angular momentum to erasure the content of the memory instead of energy. We compared the spinlabor cost for a finite reservoir to its infinite reservoir counterpart. We found that if the reservoir size is relatively small, the reservoir has a limited capacity to erase information, and correspondingly, the spinlabor cost is also low. The rate at which the finite reservoir converges to the infinite reservoir depends on the initial reservoir spin polarization $\alpha_{\rm i}$. When $\alpha_{\rm i}$ is close to 0 (i.e. corresponding to low spin temperature),  the reservoir erases the information of the memory more efficiently since the proportion of the number of reservoir particles in the spin-down compared to those in the spin-up state is larger. The average spinlabor cost per bit erased was found to be always lower for larger reservoir sizes. However, as $\alpha_{\rm i}$ increases (i.e. as the spin temperature rises) the spinlabor cost increases.

It was of particular importance and interest to calculate the reservoir size $N$ required to produce spinlabor distributions ($\mathcal{P}^{\rm fin}_N(n)$) that are similar to those ($\mathcal{P}^{\rm inf}_\infty(n)$) of their infinite reservoir counterparts. For this we used the Jensen-Shannon Divergence (JSD) \cite{Lin1991} which is a measure of how close two probability distributions are to each other.
Under the condition that the JSD is bounded by $\mbox{JSD}(\mathcal{P}^{\rm fin}_N||\mathcal{P}^{\rm inf}_\infty) \leq 0.005$ we calculated the reservoir sizes and their corresponding  probability $p_{\uparrow,\rm f}$ of the memory being in the spin-up state at the end of the erasure process. Here we found that $p_{\uparrow,\rm f}$ converges to 0 faster as $\alpha_{\rm i}$ approaches 0. In an effort to provide a more accurate account of the full cost of erasure we analyzed the cost of resetting the ancilla spin particles used in the process back their initial state, and we investigated the reuse of finite reservoirs. Our findings show that for the worse case scenario the cost of resetting the ancillas to their initial state can almost double the total cost of the entire erasure process. We also examined the effect of reusing the same finite reservoir for the erasure of additional memory spins. In contrast to the situation for an infinite reservoir which performs erasure in exactly the same way, in principle, regardless of how many times it is reused, we found that the ability to erase using a finite reservoir reduced each time it is reused.  This demonstrated the deleterious effects of the spinlabor cost of erasure that we explored elsewhere \cite{Croucher2021}. 

\section{Acknowledgments}
This research was supported by the ARC Linkage Grant No.LP180100096 and the Lockheed Martin Corporation. We acknowledge the traditional owners of the land on which this work was undertaken at Griffith University, the Yuggera people.

\newpage
\newpage
\appendix
\begin{widetext}

\section{Another form of the recurrence relation \label{sec:rec_sol}}
Here we would like to provide another form of the recurrence relation \eq{eqn:res_mem_prob_up} and its initial condition \eq{eqn:res_mem_prob_up m=0}, reproduced here as
\bea 
P_{m}(n,1) &=& T_{m}(n,1)[P_{m-1}(n,1) + P_{m-1}(n+m+1,0)] \label{eqn:res_mem_prob_up_app}
\eea
for $0 < m \le N-1$ and
\bea
P_{0}(n,1) &=& T_{0}(n,1) [P_\up(n)p_\up +  P_\up(n+1)p_\dn]. \label{eqn:res_mem_prob_up_app initial condt.} 
\eea
We divide both sides of \eq{eqn:res_mem_prob_up_app} by $ \prod_{k=0}^{m}T_{k}(n,1)$ and \eq{eqn:res_mem_prob_up_app initial condt.} by $T_{0}(n,1)$, and define 
\bea 
   A_{m} \equiv \frac{P_{m}(n,1)}{\prod_{k=0}^{m}T_{k}(n,1)},
   \qquad A_{0}\equiv P_\up(n)p_\up +  P_\up(n+1)p_\dn
\eea 
for $0 < m\le N-1$, to obtain 
\bea 
A_{m} &=& A_{m-1} + \frac{P_{m-1}(n+m+1,0)}{\prod_{k=0}^{m-1}T_{k}(n,1)} \nonumber \\
A_{m} - A_{m-1} &=& \frac{P_{m-1}(n+m+1,0)}{\prod_{k=0}^{m-1}T_{k}(n,1)}
\eea
for $0 < m \le N-1$. 
Summing both sides over $m$ from $1$ to $d$ yields
\bea 
\sum_{m=1}^{d} (A_{m} - A_{m-1}) &=&  \sum_{m=1}^{d} \frac{P_{m-1}(n+m+1,0)}{{\prod_{k=0}^{m-1}T_{k}(n,1)}}, \nonumber \\
A_{d} - A_{0} &=& \sum_{m=1}^{d} \frac{P_{m-1}(n+m+1,0)}{{\prod_{k=0}^{m-1}T_{k}(n,1)}}, \nonumber \\
A_{d} &=& A_{0} + \sum_{m=1}^{d} \frac{P_{m-1}(n+m+1,0)}{{\prod_{k=0}^{m-1}T_{k}(n,1)}}, \nonumber \\
P_{d}(n,1)  &=& \prod_{i=0}^{d}T_{i}(n,1) \left(A_{0} + \sum_{m=1}^{d} \frac{P_{m-1}(n+m+1,0)}{{\prod_{k=0}^{m-1}T_{k}(n,1)}} \right),
\nonumber \\
  &=& \prod_{i=0}^{d}T_{i}(n,1) \left(P_\up(n)p_\up +  P_\up(n+1)p_\dn + \sum_{m=1}^{d} \frac{P_{m-1}(n+m+1,0)}{{\prod_{k=0}^{m-1}T_{k}(n,1)}} \right),
\eea
where we have substituted the definition of $A_{0} = P_\up(n)p_\up +  P_\up(n+1)p_\dn$ in the last line.

\section{Average Spinlabor bound \label{sec:av spin labor bound}}
Here we present the derivation of the average spinlabor bound in \eq{eqn:bound_init} reproduced here for convenience:
\bea
\langle\mathcal{L}_{\rm s}\rangle \geq \frac{\hbar N e^{-\gamma \hbar}}{(N+1)(1+e^{-\gamma \hbar})} + \frac{\hbar}{2(N+1)}. 
\eea
As stated in the main text, the bound is derived by considering just the first term in the summation of \eq{eqn:avg_fin} which, on replacing $p_{\up,m}$ using \eq{eqn:memory_prob}, becomes
\bea
 \langle\mathcal{L}_{\rm s}\rangle = \sum_{m=1}^{N-1} \hbar \sum_{n=0}^{N} P_{m-1}(n,1) &\geq& \hbar \sum_{n=0}^{N}  P_{0}(n,1).
  \label{eqn:bound_needing_sub}
\eea
To evaluate $\sum_{n=0}^N P_{0}(n,1)$, we use the following equations 
\bea
\frac{^{N}C_{n}}{^{N}C_{n+1}+^{N}C_{n}} &=& \frac{n+1}{N+1}, \nonumber \\
\sum_{n=0}^{N} {}^{N}C_{n} e^{-\gamma n\hbar} &=& (1+e^{-\gamma\hbar})^{N}, \nonumber \\
\sum_{n=0}^{N} {}^{N}C_{n} n e^{-\gamma n\hbar} &=& N e^{-\gamma\hbar} (1+e^{-\gamma\hbar})^{N-1}, \nonumber
\eea
together with \eqs{eqn:probr}, \eqr{eqn:P_up(n)} and \eqr{eqn:res_mem_prob_up m=0}, i.e. 
\bea
P_{0}(n,1) &=& \left\{\begin{array}{l}
    T_{0}(n,1) \left[\frac{^{N}C_{n}e^{-n\gamma\hbar}}{Z_R}p_\up + \frac{^{N}C_{n+1}e^{-(n+1)\gamma\hbar}}{Z_R}p_\dn\right]\mbox{ for }0\le n \le N-1,\\
    T_{0}(N,1) \frac{e^{-N\gamma\hbar}}{Z_R}p_\up \mbox{ for } n = N,
   \end{array}\right. 
\eea
in which we replace $T_{0}(n,1)$ using \eq{eqn:T1} and substitute $p_\up=p_\dn=\frac{1}{2}$, as follows: 
\bea
\sum_{n=0}^N  P_{0}(n,1) &=& \sum_{n=0}^N \frac{^{N}C_{n}}{^{N}C_{n+1}+^{N}C_{n}}\Mycomb[N]{n}   \frac{e^{-\gamma n\hbar}}{2Z_R} +\sum_{n=0}^{N-1}  \frac{\Mycomb[N]{n}}{^{N}C_{n+1}+^{N}C_{n}}   \Mycomb[N]{n+1} \frac{e^{-(n+1)\gamma n\hbar}}{2Z_R}\nonumber \\ 
&=& \sum_{n=0}^N \Mycomb[N]{n}  \left(\frac{n+1}{N+1}\right) \frac{e^{-\gamma n\hbar}}{2Z_R} +\sum_{n=1}^{N}  \frac{\Mycomb[N]{n-1}}{^{N}C_{n}+^{N}C_{n-1}}   \Mycomb[N]{n} \frac{e^{-\gamma n\hbar}}{2Z_R}\nonumber \\ 
&=& \sum_{n=0}^N \Mycomb[N]{n}  \left(\frac{n+1}{N+1}\right) \frac{e^{-\gamma n\hbar}}{2Z_R} +\sum_{n=0}^{N}  \Mycomb[N]{n}   \left(\frac{n}{N+1}\right) \frac{e^{-\gamma n\hbar}}{2Z_R}\nonumber \\ 
&=& \sum_{n=0}^N \Mycomb[N]{n}  \left(\frac{n}{N+1}\right) \frac{e^{-\gamma n\hbar}}{Z_R} + \sum_{n=0}^{N}  \Mycomb[N]{n}   \left(\frac{1}{N+1}\right) \frac{e^{-\gamma n\hbar}}{2Z_R}\nonumber \\ 
&=&  \frac{ N e^{-\gamma \hbar}}{(N+1)(1+e^{-\gamma \hbar})} + \frac{1}{2(N+1)},
\eea
where $Z_R$ is defined in \eq{eqn:probr}. 
So \eq{eqn:bound_needing_sub} becomes
\bea
\langle\mathcal{L}_{\rm s}\rangle \geq \frac{\hbar N e^{-\gamma \hbar}}{(N+1)(1+e^{-\gamma \hbar})} + \frac{\hbar}{2(N+1)}
\eea
as required.

\section{Values of Fig. 11 and Fig 12 \label{sec:Table}}

\begin{table}[H]
\begin{center}
    \begin{filecontents*}{table_data.csv}
$\alpha_{i}$,Reservoir Size,$p_{\uparrow,f}$,$\alpha_{i}$,Reservoir Size,$p_{\uparrow,f}$
0.01,21,4.19e-13,0.21,26,0.0014
0.02,17,6.60e-09,0.22,29,0.0012
0.03,14,1.30e-06,0.23,31,0.0013
0.04,13,1.20e-05,0.24,34,0.0012
0.05,13,3.26e-05,0.25,37,0.0012
0.06,13,7.42e-05,0.26,41,0.0011
0.07,13,0.0001,0.27,45,0.0011
0.08,13,0.0003,0.28,50,0.0010
0.09,13,0.0005,0.29,55,0.0011
0.10,13,0.0008,0.30,62,0.0009
0.11,14,0.0008,0.31,70,0.0009
0.12,15,0.0008,0.32,79,0.0008
0.13,15,0.0012,0.33,90,0.0008
0.14,16,0.0012,0.34,103,0.0007
0.15,17,0.0013,0.35,119,0.0006
0.16,18,0.0014,0.36,140,0.0006
0.17,20,0.0012,0.37,165,0.0005
0.18,21,0.0013,0.38,198,0.0004
0.19,23,0.0012,0.39,241,0.0004
0.20,24,0.0015,0.40,299,0.0003
\end{filecontents*}
\csvreader[no head,tabular=|c|c|c||c|c|c|,
table head=\hline,late after line=\\\hline]{table_data.csv}
    {1=\one, 2=\two, 3=\three, 4=\four, 5=\five, 6=\six}
    {\one & \two & \three & \four & \five & \six}
    \caption{Data used by Fig. \ref{fig:similarity_between_work_prob} and Fig. \ref{fig:similarity_between_erasure}.}\label{tab:table_data} % 
    %\centering
\end{center}
\end{table}

\end{widetext}

\bibliography{bib1}

\end{document}